# Intercomparison exercise on Monte Carlo simulations of electron spectra and energy depositions by a single gold nanoparticle under X-ray irradiation

Wei Bo Li[1], Hans Rabus[2], Carmen Villagrasa[3], Jan Schuemann[4]

[1] Bundesamt für Strahlenschutz, Neuherberg, Germany

[2] Physikalisch-Technische Bundesanstalt, Berlin, Germany

[3] Institut de Radioprotection et de Sûreté Nucléaire, Fontenay-Aux-Roses, France

[4] Massachusetts General Hospital, Department of Radiation Oncology, Boston, USA

Editor: Filip Vanhavere[1,2]

[1] Nuclear Research Centre (SCK.CEN), Belgium

[2] Chair, EURADOS e.V.

**List of contributors to the report (affiliations at the time of the exercise)**

| | |
|---|---|
| Michael Beuve | Institut de Physique des 2 Infinis, Université Claude Bernard Lyon 1, Villeurbanne, France |
| Marion Bug | Physikalisch-Technische Bundesanstalt Braunschweig, Germany |
| Yi Zheng Chen | Department of Engineering Physics, Tsinghua University, Beijing, China |
| Salvatore Di Maria | Instituto Superior Técnico, Universidade de Lisboa, Bobadela, Portugal |
| Werner Friedland | Institute of Radiation Medicine, Helmholtz Zentrum München - German Research Center for Environmental Health, Neuherberg, Germany |
| Liset de la Fuente Rosales | Physikalisch-Technische Bundesanstalt, Braunschweig, Germany |
| Bernd Heide | Karlsruhe Institute of Technology, Karlsruhe, Germany |
| Philine Hepperle | Physikalisch-Technische Bundesanstalt, Braunschweig, Germany |
| Nora Hocine | Institut de Radioprotection et de Sûreté Nucléaire, Fontenay-Aux-Roses, France |
| Alexander Klapproth | Institute of Radiation Medicine, Helmholtz Zentrum München – German Research Center for Environmental Health, Neuherberg, Germany |
| | TranslaTUM, Klinikum rechts der Isar, Technische Universität München, Munich, Germany |
| Chun Yan Li | Department of Engineering Physics, Tsinghua University, Beijing, China |
| Jun Li Li | Department of Engineering Physics, Tsinghua University, Beijing, China |
| Wei Bo Li | Institute of Radiation Medicine, Helmholtz Zentrum München - German Research Center for Environmental Health, Neuherberg, Germany |
| Heidi Nettelbeck | Physikalisch-Technische Bundesanstalt, Braunschweig, Germany |
| Floriane Poignant | Université Claude Bernard Lyon 1, Villeurbanne, France |
| Hans Rabus | Physikalisch-Technische Bundesanstalt, Braunschweig, Germany |
| Benedikt Rudek | Massachusetts General Hospital, Department of Radiation Oncology, Boston, USA |
| Jan Schuemann | Massachusetts General Hospital, Department of Radiation Oncology, Boston, USA |
| Carmen Villagrasa | Institut de Radioprotection et de Sûreté Nucléaire, Fontenay-Aux-Roses, France |

**Imprint**

Issued by:
European Radiation Dosimetry e. V.
Postfach 1129
85758 Neuherberg
Germany
office@eurados.org
www.eurados.org

**Liability Disclaimer**

# Contents

# Abstract

In vitro and in vivo experiments showed that gold nanoparticles (GNPs) enhance radiation effects to tumours in mice and demonstrated a high potential of GNPs for clinical application mostly as a contrast agent in kV imaging but so far for MV radiotherapy treatment using GNPs or other NP is illusive or of marginal. This enhanced biological effectiveness is often attributed to the enhanced energy deposition in the vicinity of the GNP by secondary electrons around GNPs, particularly by short-ranged Auger electrons. However, this energy deposition is difficult to measure. Energy deposition or dose is not the only metric that may be important for radiobiological response of cells to ionizing radiation in the presence of GNPs. Computational approaches, such as Monte Carlo (MC) radiation transport simulations, are used to estimate the dosimetric effects of GNPs, where results differing by orders of magnitudes have been reported by different investigators. This has motivated an intercomparison exercise, which was conducted as a joint activity of EURADOS Working Groups 6 "Computational Dosimetry" and 7 "Internal Dosimetry". The aim of this exercise was to determine the extent of such discrepancies between the results obtained by different researchers and different codes in a very simple simulation setup.

Several individual EURADOS associate members and two code developer groups from outside Europe participated in this exercise applying seven different MC codes to perform the simulations of a simple defined geometry set-up of one single GNP irradiated in water by kilo-voltage X-rays. Two GNP diameters of 50 nm and 100 nm of were considered and two photon spectra as generated by X-ray tubes operated at 50 kV and 100 kV peak voltages. The geometry set-up and X-ray spectra were provided by the EURADOS task group. The participants were asked to determine for each combination of GNP size and X-ray spectrum the dose enhancement ratio (DER) of 10 nm-thick water shells up to 1000 nm and 1 μm-thick water shells up to 50 μm around the GNP. Furthermore, the electron spectra emitted from the GNP and the energy depositions in water shells around it were also to be reported.

This EURADOS report summarizes the motivation and background for the exercise, the tasks to be solved, the codes used, the results reported by the participants, the consistency checks applied in their evaluation and a best estimates and uncertainty bands derived from the final results for the energy spectra of emitted electrons and the energy imparted in the vicinity of the GNP.

# 1. Introduction

In the therapeutic application of ionizing radiation for cancer treatment, there is a continuous endeavor to reduce the radiation dose delivered to normal tissues of the patient, while keeping or even enhancing the therapeutic dose to the tumor, so that the risk of developing toxic side effects can be reduced (Halperin *et al.*, 2013). Among many novel techniques, a very promising approach to achieve this aim is to use radiosensitizers, which are defined as substances that make tumor cells more sensitive to radiation-induced cell killing, without changing the sensitivity of cells in healthy tissues. In this context, nanoparticles made of high-Z materials have been investigated as potential radiosensitizers (Kuncic and Lacombe, 2018; Schuemann *et al.*, 2020; Schuemann *et al.*, 2016). Due to their presumed biocompatibility, their strong photoelectric absorption coefficient, and the emission of Auger and Coster–Kronig (C-K) electrons, gold nanoparticles (GNPs) (Z=79) have been extensively investigated for several years as possible agents for a selective amplification of the radiation dose in tumors (Bergs *et al.*, 2015). This concept is known as "gold nanoparticle assisted radiation therapy" (GNRT) (Cooper *et al.*, 2014; Zygmanski and Sajo, 2016; Lin *et al.*, 2015; Lin *et al.*, 2014; Sung *et al.*, 2018; Gadoue *et al.*, 2018; Her *et al.*, 2017; Cui *et al.*, 2017; Mesbahi, 2010; Dorsey *et al.*, 2013).

The first successful experiment using GNPs to increase the radiosensitivity of tumors in mice irradiated by x-rays (Hainfeld *et al.*, 2004) stimulated extended investigations, by experiments (Chithrani and Chan, 2007; Chithrani *et al.*, 2010a; Chithrani *et al.*, 2006; Chithrani *et al.*, 2010b; Yang *et al.*, 2014; Chattopadhyay *et al.*, 2013; Chattopadhyay *et al.*, 2010) and computational simulations, into the effects of GNPs when using different types of radiation such as kilovoltage x-rays (Cho, 2005; Jones *et al.*, 2010; Leung *et al.*, 2011; Lechtman *et al.*, 2013; Douglass *et al.*, 2013; Zygmanski *et al.*, 2013; Li *et al.*, 2014; Xie *et al.*, 2015; Carter *et al.*, 2007; Cho *et al.*, 2009; Lechtman *et al.*, 2011; McMahon *et al.*, 2008), megavoltage x-rays, protons and heavy ions (Jain *et al.*, 2011; Kim *et al.*, 2012; Kaur *et al.*, 2013) and radionuclide like $^{125}$I (Brivio *et al.*, 2017). The reported results suggest that the sensitization of cells by GNPs strongly depends on the particle type and energy spectrum of the incident radiation and that it may result from an enhanced energy deposition in the vicinity, up to micrometers, but mostly within several tens of nanometers around the GNPs. A wide range of results were reported in these studies, which means that the results are highly dependent on how the MC simulation is performed and which metrics are evaluated. For many approximate approaches, there were many assumptions and shortcuts that led to ambiguous results that were difficult to compare with other approaches.

Since the enhancement is extremely localized, a selective uptake of GNPs into or around tumor cells is required. A high concentration of GNPs in the cancer cells or around them can be achieved by targeting antibodies. The antibody cmHsp70.1 can, for instance, be conjugated with GNPs and accumulate them into breast cancer cells (Stangl *et al.*, 2011; Gehrmann *et al.*, 2015). When irradiating cancer cells loaded with GNPs by x-rays, the enhanced radiation dose can destroy the cell membrane (Fink and Cookson, 2005), mitochondria (McMahon *et al.*, 2017) and even the DNA (Lomax *et al.*, 2013) and consequently kill the cancer cells. As this effect occurs near the GNPs, the surrounding healthy cells and tissues are preserved. Many factors, e. g. the shape, size and uptake of GNPs as well as the type of cell line, influence the biodistribution of GNPs inside cells and subsequently affect the dose enhancement and biological outcome, e. g. cell survival fraction, DNA damages and mice survival rate (Chithrani *et al.*, 2006; Chithrani and Chan, 2007; Jain *et al.*, 2011; Cho *et al.*, 2011). All these different biological endpoints will be altered in the presence of GNPs and thus the physical radiation doses must be reliably assessed to establish the dose–effect relationship in order to facilitate the use of GNPs in this new potential GNRT modality. (However, as was pointed out by Zygmanski *et al* (2023), GNPs also have biochemical effects due to their interaction with the cellular environment, where some interaction channels can be triggered by irradiation.)

A second issue with the localization of the enhanced energy deposition is that it cannot be directly measured with present technology, such that its determination requires Monte Carlo (MC) or other numerical simulations such as deterministic radiation transport calculations, where very different results have been reported in the literature (Vlastou *et al* 2020, Moradi *et al* 2021). In some cases, even when the same code was used, various options for the physical models and cross sections in the code applied by different modelers lead to different results.

Therefore, in 2013 during the EURADOS annual meeting in Barcelona, members of EURADOS Working Groups 6 “Computational Dosimetry” and 7 “Internal Dosimetry” organized a joint meeting on the determination of dose enhancement by gold nanoparticles in radiotherapy. As a result, a EURADOS exercise on simulations of single GNP irradiated by kilovoltage X-rays was proposed and conducted during 2015 to 2019. The aim of this exercise was to determine the spread in the results obtained by different participants using different MC codes.

In the exercise, electron energy spectra and absorbed dose ratios in the immediate vicinity (nanometer ranges) and at larger distances (micrometer ranges) from a GNP were to be simulated. The exercise was performed by different research groups involved in the EURADOS network and other MC simulation labs from the USA and China by applying a simple geometry, where one single GNP was irradiated by pre-defined x-ray spectra. The participants used different MC codes, namely Geant4/Geant4-DNA, TOPAS/TOPAS-nBio, PENELOPE, PARTRAC, NASIC, MDM and MCNP6

Preliminary results of the exercise were presented at the following conferences: (1) International Conference ARGENT - Advanced Radiotherapy, Generated by Exploiting Nanoprocesses and Technologies, January 22-24, 2018 in , Gif-sur-Yvette, France; (2) 3rd Geant4 International User Conference "At the Physics-Medicine-Biology frontier", October 29-31, 2018 in Bordeaux, France; (3) The 3rd International Conference on Dosimetry and its Applications (ICDA-3), May 27-31, 2019 in Lisbon, Portugal. Based on an abstract of the preliminary status of the exercise analysis in the proceedings of the 3rd Geant4 International User Conference, an extended manuscript was produced and published in Physica Medica (Li *et al.*, 2020a). The large diversity in the electron spectra caused a re-evaluation of the published results. Some inconsistencies on the normalization of electron spectra, the input of X-ray spectra and the implementation of defined geometry set-up were identified (Rabus *et al.*, 2021a), and corrected results were published as a corrigendum in Physica Medica (Li *et al.*, 2020b).

In this report, a summary of the exercise is presented along with the methods used for consistency check and the final results. Some remaining inconsistency issues and experience learned from the exercise are also discussed.

# 2. Description of the Exercise

## 2.1 Simulation set-up

A single spherical GNP was positioned at the center of the simulation tracking volume consisting of liquid water. In this exercise, the GNP was assumed as pure gold (without coating and without any conjugation with an antibody). Two particle sizes in diameter of ∅ = 50 nm and ∅ = 100 nm were used in the simulations. This simple assumption of a pure gold particle facilitates the comparison of simulation results as it avoids effects of energy absorption in the coating and the antibody.

Figure 1 shows the set-up of the simulation geometry. A single GNP with a diameter of 50 nm or 100 nm is located in liquid water. Parallel x-rays, produced by an x-ray tube, are sampled from a planar circular area with a diameter of 10 nm larger than the diameter of the GNP and irradiate the GNP along the z-axis in a right-handed Cartesian reference frame. The distance between the center of the planar circular x-ray source and the center of the single GNP is 100 µm. Energy deposition is scored in concentric spherical shells of thickness d around the GNP.

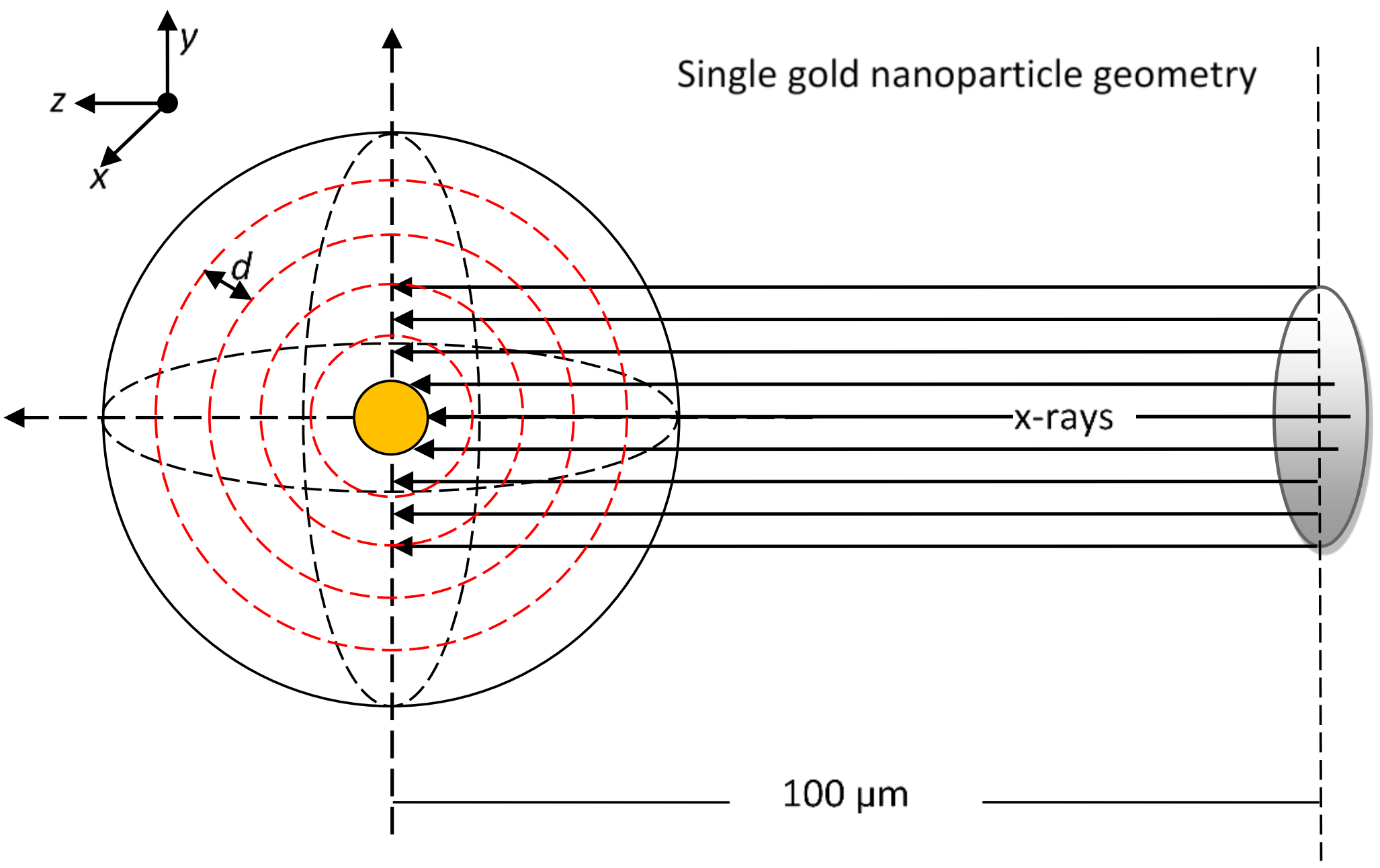

Figure 1: Geometry set-up of the simulation exercise (not to scale). Reproduced with permission of the copyright owner Elsevier (license no. 5531310292244) from Li at al., *Intercomparison of dose enhancement ratio and secondary electron spectra for gold nanoparticles irradiated by X-rays calculated using multiple Monte Carlo simulation codes*, Physica Medica 69, 147-163 (2020). A single spherical gold nanoparticle (orange circle) surrounded by liquid water is irradiated by a parallel beam of X-rays of energy spectra as shown in Figure 2. The photon beam is emitted by a circular source located at 100 µm distance from the nanoparticle and has a diameter 10 nm larger than the diameter of the nanoparticle. Nanoparticle diameters 50 nm or 100 nm were considered. Energy imparted was scored in spherical shells around the GNP, where the surface of the largest sphere was at 50 µm radial distance from the GNP surface.

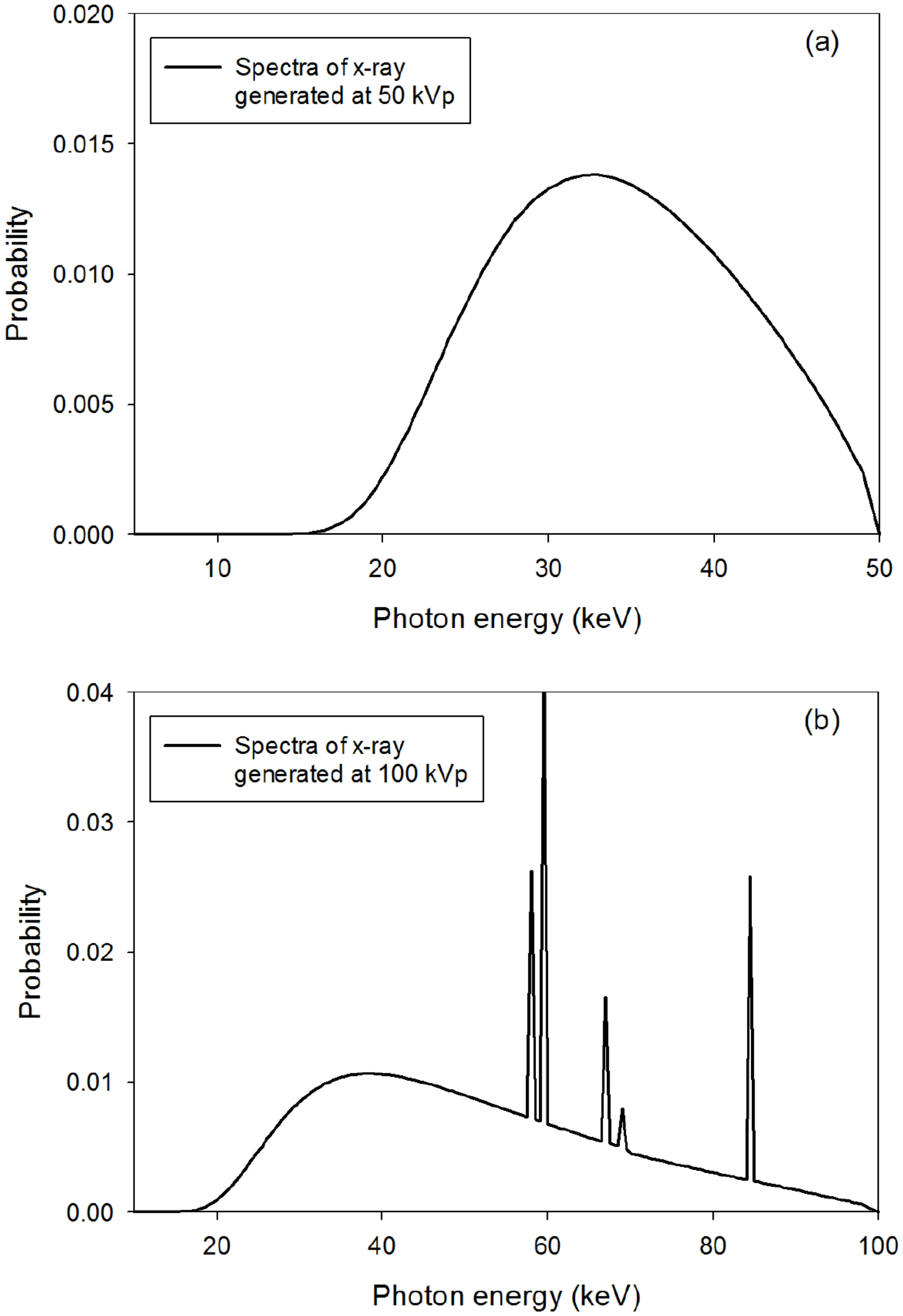

Figure 2: Histogram of the probability of emission of photons in 500 eV bins for the 50 kVp (a) and 100 kVp (b) X-ray spectra used in the exercise. (The peak at about 84.5 keV in the 100 kVp spectrum is physically not plausible and presumably due to a shifted decimal place as the value is about a factor of ten higher than the mean of the neighbouring data points.)

The x-ray spectra provided to the participants were calculated by the program SpekCalc (Poludniowski *et al.*, 2009) simulating an x-ray tube with a tungsten target. The parameters used to produce the x-ray spectra are as follows:

- Peak voltage (kVp): 50 and 100
- Energy bin (keV): 0.5
- Angle theta (degree): 20
- Air thickness (mm): 470
- Beryllium thickness (mm): 0.8
- Aluminium thickness (mm): 3.9
- Nf: 0.68
- P: 0.33

The model parameters 'Nf' and 'P' in the GUI interface take the default values of 0.68 and 0.33 (Poludniowski, 2007). The former normalizes the overall fluence and can be used to match the output prediction to that of a particular x-ray tube, if desired. The latter is the ratio of the number of characteristic x-ray photons produced via electron impact ionization to the number of photons generated by bremsstrahlung interaction with the atomic nucleus; this ratio should not be changed without justification (Poludniowski *et al.*, 2009). The resulting spectral photon energy fluence of these two x-ray spectra, 50 kVp and 100 kVp is shown in Figure 2.

It is understood that the simple irradiation geometry shown in Figure 1 is far from any realistic clinical scenario. One single photon microbeam cannot be used in the clinical environment and targeting a single nanoparticle with such a beam is unfeasible. Furthermore, this irradiation setup implies lateral particle disequilibrium. However, the purpose of the exercise was not to draw conclusions about realistic scenarios. Like in many other EURADOS intercomparisons in computational dosimetry, the intent was rather to assess the spread of results that are obtained when different users try to solve the same problem. The simulation setup was intentionally chosen this simplistic in order to exclude the additional error sources from multiscale particle transport simulation that may be the origin of most of the confusion and conflicting results in literature.

## 2.2 Results to be reported

The X-ray spectra simulated for an X-ray tube with a tungsten target and 50 kV and 100 kV peak acceleration voltages shown in Figure 2 were provided to the participants as cumulative spectra to be read by each MC code. The cut-off energies for transport of photons and electrons were selected by modelers according to the cross sections provided in their MC codes. The geometry setup of GNPs and x-ray sources defined in this exercise were implemented in the simulations by the participants.

The main quantity to be determined by participants was the ratio of the average energy deposited within spherical water shells resulting from x-ray irradiation with and without GNP in the center. The liquid water surrounding the GNP was divided into concentric shells of equal thickness as shown in Figure 1. Starting from the surface of the GNP, 50 water shells with an equal thickness of d = 1 µm were set as sensitive target volumes to mimic cellular targets. The energy deposition was also to be scored with finer resolution in 100 concentric water shells with an equal thickness of d = 10 nm starting from the surface of GNP. In addition to the energy deposition in the water shells, the energy fluence of secondary electrons and Auger electrons originating from the GNP were to be scored for further analysis.

The ratio of the average energies deposited in the target volumes with and without GNP was originally referred to in the exercise as the "dose enhancement ratio". Since the simplified geometry setup does not ensure lateral secondary charged particle equilibrium, the energy imparted in these targets is not representing the absorbed dose under realistic irradiation conditions. Therefore, in this report this quantity is called deposited-energy ratio (DER).

## 2.3 Monte Carlo codes used within the exercise

Table 1 lists the codes used in the exercise along with respective reference information.

Table 1: Monte Carlo codes employed in the frame of the exercise.

| Code | References |
|---|---|
| Geant4/Geant4-DNA | (Incerti *et al.*, 2010; Bernal *et al.*, 2015; Incerti *et al.*, 2016; Incerti *et al.*, 2018) |
| MCNP6 | (Goorley *et al.*, 2012) |
| MDM | (Gervais *et al.*, 2006) |
| NASIC | (Li *et al.*, 2015) |
| PARTRAC | (Friedland *et al.*, 2011) |
| PENELOPE | (Salvat, 2015; Salvat *et al.*, 2011; Salvat, 2019) |
| TOPAS-nBio | (Perl *et al.*, 2012; Schuemann *et al.*, 2019) |

Table 2: Details of the Monte Carlo codes and options used by the participants for their simulations. Reproduced with permission of the copyright owner Elsevier (license no. 5531310292244) from Li at al., *Intercomparison of dose enhancement ratio and secondary electron spectra for gold nanoparticles irradiated by X-rays calculated using multiple Monte Carlo simulation codes*, Physica Medica 69, 147-163 (2020).

| Participant ID | Code name and version | Code option, processes considered and related data libraries[1] | | | | electron simulation mode in Au[2] | Cut-off energy | | Number of primary particles |
|---|---|---|---|---|---|---|---|---|---|
| | | photons (Au & $H_2O$) | electrons in $H_2O$ | electrons in Au | deexcitation | | photon | electron | |
| G4/DNA#1 | Geant4/DNA 2016 (Default option and option 7) | Coherent and incoherent scattering, photoelectric absorption, pair production (Geant4 EM default) | Inner-shell impact ionization, excitation, attachment (Geant4-DNA default option) | Livermore processes and multiple scattering | X-ray fluorescence and Auger electrons for KLM shells | Condensed history (Livermore)) | 50 eV | 10 eV | $10^8$ |
| G4/DNA#2 | Geant4/DNA 10.0.5 | Coherent and incoherent scattering, photoelectric absorption, pair production (Geant4 EM default) | Geant4-DNA default physics (inner-shell impact ionization, excitation, attachment) | Livermore processes | Particle induced X-ray emission; complete Auger deexcitation | Condensed history | 990 eV | 10 eV | $10^7$ |

[1] Unless stated differently in the table, the codes used cross section data from the Evaluated Photon Data Library (EPDL) and the Evaluated Electron Data Library (EEDL) and atomic relaxation data from the Evaluated Atomic Data Library (EADL).

[2] All participants used track structure simulation in water.

| Participant ID | Code name and version | Code option, processes considered and related data libraries[1] | | | | electron simulation mode in Au[2] | Cut-off energy | | Number of primary particles |
|---|---|---|---|---|---|---|---|---|---|
| | | photons (Au & $H_2O$) | electrons in $H_2O$ | electrons in Au | deexcitation | | photon | electron | |
| G4/DNA#3 | Geant4/DNA 10.4.2 | Coherent and incoherent scattering, photoelectric absorption, pair production (Geant4 EM default) | Geant4-DNA default option (inner-shell impact ionization, excitation, attachment) | Geant4 EM standard physics option 4 | Particle induced x-ray emission, Auger electrons from K-, L- and M-shells and Auger cascades (EADL) | condensed history | 10 eV | 10 eV | $10^9$ |
| MCNP6 | MCNP6.1 2013 | coherent and incoherent scattering, photoelectric absorption, and electron/positron pair production (ENDF/B VI.8) | atomic excitation, electron elastic scattering, subshell electron impact ionization, and bremsstrahlung (ENDF/B VI.8) | atomic excitation, electron elastic scattering, subshell electron impact ionization, and bremsstrahlung (ENDF/B VI.8) | Relaxation considering 29 subshells and almost 3,000 transitions | single-event method | 1 keV | 50 eV | $10^8$ |
| MDM | MDM 2006 (water) 2019 (gold) | only photoabsorption and incoherent scattering (NIST XCOM) | Elastic scattering, BEB Excitations (2 modes), vibrations (9 modes). | Inelastic: Plasmon excitation BEB Elastic: ELSEPA Phonons | Full Auger electronic cascade from K-shell to valence band | Track structure | No explicit tracking of photons | Water: 7 eV Gold: 10 eV | $7.1\times10^8$ for 50 nm GNP, $2.4\times10^9$ for 100 nm GNP |

| Participant ID | Code name and version | Code option, processes considered and related data libraries[1] | | | | electron simulation mode in Au[2] | Cut-off energy | | Number of primary particles |
|---|---|---|---|---|---|---|---|---|---|
| | | photons (Au & $H_2O$) | electrons in $H_2O$ | electrons in Au | deexcitation | | photon | electron | |
| NASIC | NASIC 2018 | Coherent and incoherent scattering, photoelectric absorption, pair production (Geant 4 EM default) | ionization and excitation (NASIC cross section library) elastic scattering, attachment, vibrational excitation (Geant4-DNA default) | Multiple scattering, ionization, bremsstrahlung (Geant4 default) | X-ray fluorescence and Auger electrons | Condensed history | 10 eV | 10 eV | $3\times10^8$ |
| PARTRAC | PARTRAC 2015 | Coherent and incoherent scattering, photoelectric absorption | Ionization: 5 shells, Excitation: 5 levels (Dingfelder[3]) | Ionization: 13 shells and subshells, Excitation: 1 level (Hantke[4]) | All transitions from EADL library | Track structure | 100 eV | ($H_20$):10 eV (Au): 100 eV | $>10^9$ |
| PENELOPE#1 | PENELOPE 2011 and 2018 (for revised results) | Coherent and incoherent scattering, photoelectric absorption, pair production | Inner-shell impact ionization (Penelope cross section library) | Inner-shell impact ionization (Penelope cross section library) | X-ray fluorescence and Auger electrons from K N shell vacancies | Track structure | 50 eV | 50 eV | $10^7$ |

[3] Dingfelder et al. 1998

[4] Unpublished data

| Participant ID | Code name and version | Code option, processes considered and related data libraries[1] | | | | electron simulation mode in Au[2] | Cut-off energy | | Number of primary particles |
|---|---|---|---|---|---|---|---|---|---|
| | | photons (Au & $H_2O$) | electrons in $H_2O$ | electrons in Au | deexcitation | | photon | electron | |
| PENELOPE#2 | PENELOPE 2014 | Coherent and incoherent scattering, photoelectric absorption, pair production | Inner-shell impact ionization (Penelope cross section library) | Inner-shell impact ionization (Penelope cross section library) | X-ray fluorescence and Auger electrons from K to N shell vacancies | Track structure | 50 eV | 50 eV | $10^7$ |
| TOPAS | TOPAS 3.1.p3 | Coherent and incoherent scattering, photoelectric absorption, pair production | Inner-shell impact ionization, excitation, attachment, bremsstrahlung | Inner-shell impact ionization, excitation, bremsstrahlung | X-ray fluorescence and Auger electrons for KLM shells | Condensed history (Livermore) | 50 eV | 10 eV | $5.1\times10^7$ for 50 kVp, $3.4\times10^7$ for 100 kVp |

## 2.4 Solutions provided by the participants

Table 3 and Table 4 give an overview of the solutions delivered by the 10 participants in the exercise.

Table 3: Overview of the original solutions delivered by the participants for the different parts of the exercise. Full details on the usage of the codes can be found in Table 2. Cases (a) 50 kVp, 50 nm GNP, (b) 50 kVp, 100 nm GNP, (c) 100 kVp, 50 nm GNP, (d) 100 nm GNP, 100 kVp.

| Participant ID | Code used | Solutions delivered | |
|---|---|---|---|
| | | DER | Electron spectra; bin size |
| G4/DNA#1 | GEANT4-DNA | all | all; 5 eV |
| G4/DNA#2 | GEANT4-DNA | - | (a) and (b); 100/decade |
| G4/DNA#3 | GEANT4-DNA | all | all; 5 eV |
| MCMP6 | MCNP6 | all | all; 50 eV |
| MDM | MDM | all | all; 50 eV |
| NASIC | NASIC | all | all; 10 eV |
| PARTRAC | PARTRAC | all | all; 100/decade |
| PENELOPE#1 | PENELOPE 2011 | all | all; 50 eV for 50 kVp, 100 eV for 100 kVp |
| PENELOPE#2 | PENELOPE 2014 | all | not delivered |
| TOPAS | TOPAS-nBio | all | all; 10 eV |

Table 4: Revised solutions delivered by some participants in the course of the exercise or after the analysis presented in chapter 3. Cases (a) 50 kVp, 50 nm GNP, (b) 50 kVp, 100 nm GNP, (c) 100 kVp, 50 nm GNP, (d) 100 nm GNP, 100 kVp.

| Participant ID | Code used | Solutions delivered | |
|---|---|---|---|
| | | DER | Electron spectra; bin size |
| G4/DNA#2 | GEANT4-DNA | - | (a); 100/decade |
| MCNP6 | MCNP6 | - | (a); 50 eV |
| MDM | MDM | - | all; 50 eV |
| PENELOPE#1 | PENELOPE 2018 | all | (a) and (c) 50 eV, (b) and (d) 100 eV |

# 3. Consistency checks

The submitted results of the participants were passed through a number of plausibility and consistency checks. The procedures allowed, in some cases, identifying the origin of deviations (such as improper implementation of the exercise description) and a means for correcting the results. In essence, these plausibility and consistency checks fall into two categories: Some are related to prior knowledge on the physical interaction processes and the resulting expected relative shape of emitted electron spectra and the radial energy deposition around the GNP. The other category is based on the fundamental principle of energy conservation which allows testing the data on an absolute scale. For instance, the energy deposited by electrons produced in a photon interaction cannot exceed the incident photon energy and must be in the order of the energy transferred in the interaction.

## 3.1 Plausibility of the dependence of electron energy spectra with GNP size and photon spectrum

### *3.1.1 Physical background: interactions of X-rays and electrons with gold and liquid water*

Photons interact with matter by three main processes transferring energy: photoelectric absorption, incoherent or Compton scattering and electron-positron pair production. A fourth process, Rayleigh or coherent scattering, is important at energies below 100 keV especially for high-Z materials, such as gold (for which it contributes up to 15% to the total interaction cross section in this energy range). This is an elastic interaction and impacts only the direction of propagation of the photon. The cross section for photoelectric absorption strongly depends on the nuclear charge number Z of the atom and the photon energy *E*, like $\sigma \propto (Z/E)^n$, with n≅3-4. For high-Z atoms (Z=60-80), photoelectric absorption is the dominant interaction for photons of energies up to around 500 keV, for water it is the dominant interaction up to about 28 keV photon energy. Compton scattering becomes the dominant interaction process with high-Z atoms as the photon energy becomes greater than 500 keV; for water it is dominant from about 28 keV. Electron-positron pair production can only occur for photon energies larger than 1.022 MeV. Thus, for the photon spectra used in the exercise, only photoabsorption and elastic scattering are relevant for photons interacting with gold atoms, whereas for water only photoabsorption and Compton scattering contribute.

Calculated cross sections for photon interactions are available in the form of mass attenuation coefficients from the XCOM database of NIST (Berger *et al.*, 2010). Figure 3(a) shows the total mass attenuation coefficients for the interaction of photons with liquid water and gold in the energy range considered in the exercise. Figure 3(b) shows the mean free path of a photon in water and gold, which can be directly calculated from the mass attenuation coefficient shown in Figure 3(a). Figure 3(c) and (d) show the resulting probabilities of photon interacting in the water sphere or the GNP for the irradiation geometry used in the exercise and the (c) 50 kVp and (d) 100 kVp spectra, respectively. The blue lines in Figure 3(c) and (d) show the probability of a photon emitted from the source to interact with water in a sphere of 50 µm radius around the GNP position (as was used for scoring energy deposition). The solid and dashed orange lines show the probability that such a photon interacts in a GNP of 100 nm and 50 nm diameter, respectively.

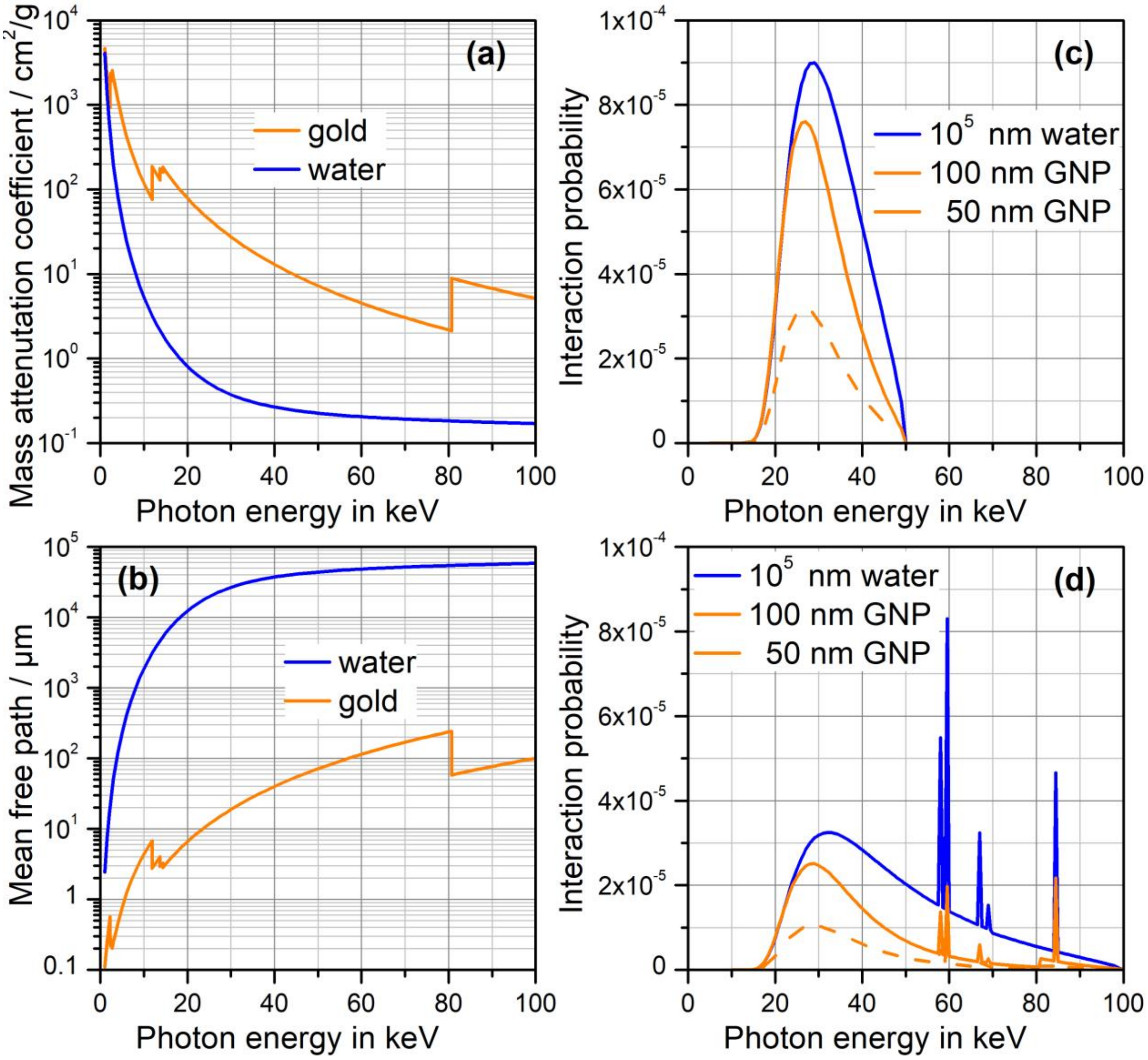

Figure 3: (a) Total mass attenuation coefficients for photon interaction with liquid water and gold. (b) mean free path for photon propagation in gold and liquid water. (c) and (d) Probability for a photon emitted from the source in the simulations to interact in a water sphere of $10^5$ nm (i.e., 100 µm) diameter or in the 100 nm GNP or 50 nm GNP for (c) the 50 kVp spectrum and (d) the 100 kVp spectrum.

Figure 3(c) and (d) illustrate that the narrow-beam irradiation geometry used in the exercise brings the probability for a photon interaction to occur in the GNP to the same order of magnitude as the probability of a photon interacting when traversing the water sphere of 100 µm (i.e., $10^5$ nm) diameter used for scoring the energy deposition. (For a larger source, the probability of an emitted photon to interact with the GNP would scale inversely proportional to the source size, whereas the probability of interaction in a 100 µm slab of water would remain the same.) The interactions in water are spread out over a stretch of 100 µm, while mainly electrons from such interactions in the vicinity of the GNP contribute to the flux of electrons leaving the GNP. Therefore, this contribution is expected to be negligible.

For photoelectric absorption, the ejected electron comes predominantly from the innermost shell from which the photon can release an electron. Therefore, the majority of photon interactions occurring in the GNP (shown in Figure 3(c) and (d)) lead to the ionization of one of the L-shells. For the 100 kVp spectrum, the small proportion of interactions from photons with energies exceeding the K-shell binding energy ($E_K$=80.8 keV) mainly ionizes the gold K-shell. For the maximum photon energy of 100 keV, the K-shell photoelectron has an energy of about 19.2 keV and a range in liquid water of about 10 µm (Meesungnoen *et al.*, 2002). The maximum energy of Compton electrons produced in water by interactions of photons from the 100 kVp spectrum is also about 20 keV, The photoelectrons produced in water and photoelectrons from an L-shell ionization of gold can have kinetic energies up to about 100 keV and 88 keV, respectively, for the 100 kVp spectrum. For the 50 kVp spectrum, the maximum photoelectron energy is about 38 keV for L-shell ionization of a gold atom and almost 50 keV for water. Therefore, a large proportion of the photoelectrons produced in the irradiations of the exercise has ranges between 10 µm and 100 µm in water (Rabus *et al.* 2019).

A core-shell ionized atom de-excites by the emission of Auger electrons or fluorescence x-rays. A K-shell vacancy of gold is filled with more than 96 % probability by a radiative transition and the emission of characteristic X-ray photons with energies mainly between 68 keV and 78 keV (data from ENDF database explorer: https://www-nds.iaea.org/exfor/). As can be seen from Figure 3(b), these electrons have a range in water of several cm. Their mean free path in gold is in the order of 200 µm. Electrons emitted by Auger decay of a K-shell vacancy have energies between 52 keV and 75 keV with a range in water of 50 µm to 80 µm.

Vacancies in the $L_1$, $L_2$ and $L_3$-shell of gold are filled by radiative transitions with probabilities of about 8 %, 35 % and 32 %, respectively, where the dominant photon peaks have energies about 9.7 keV and 11 keV. Such photons have mean free paths in water of a few mm (Figure 3(b)) and, therefore, do not contribute to a local energy deposition around the GNP (like the K-shell fluorescence photons). Their mean free path in gold is in the range between 4 µm and 5 µm (Figure 3(b)) so that there is a small probability in the order of 0.5 % to 1 % of them being reabsorbed in a 50 nm GNP and 100 nm GNP, respectively. The non-radiative de-excitation of the gold $L_1$-shell is dominated by Coster-Kronig transitions producing electrons of energies in the range from 100 eV to 200 eV, with some additional Auger electron lines with energies between 2 keV and 2.5 keV and between 11 keV and 12 keV. The emitted Auger electrons from the $L_2$ and $L_3$-shell mostly have energies between 5.5 keV and 9.5 keV and a range in water of less than 1 µm up to about 2.5 µm.

For the M- and N-shells for gold, the dominant de-excitation process is Auger electron emission with a total probability of 98.5% for the M shells and almost 100% for the N shells (Perkins *et al* 1991). N-shell Auger electrons have energies mostly in the order of a few hundred eV, M shell Auger electrons have energies up to about 3 keV, where about 30% of the have energies below 500 eV. These Auger electrons deposit their energies in the immediate vicinity of the GNP and contribute mostly to a local dose enhancement effect.

### *3.1.2 Note on the adequate presentation of the emitted electron energy spectra*

It should be noted that throughout this report electron energy spectra are presented with a logarithmic *x*-axis and a linear *y*-axis showing the frequency density multiplied by the electron energy. This way of presentation has the advantage (in analogy to microdosimetric spectra) that the area under the curve is proportional to the contribution of the respective energy range to the total number of emitted electrons.

In some cases, such as Figure 4 and Figure 5 below, the data additionally have been divided by the expected number of photon interaction in a gold nanoparticle of the respective size when irradiated with a beam of photons of a spectral fluence equal to that of the primary photon spectrum. The resulting quantity is an estimate for the energy flux of outgoing electrons at the surface of a GNP that experienced a photon interaction from such a photon energy spectrum.

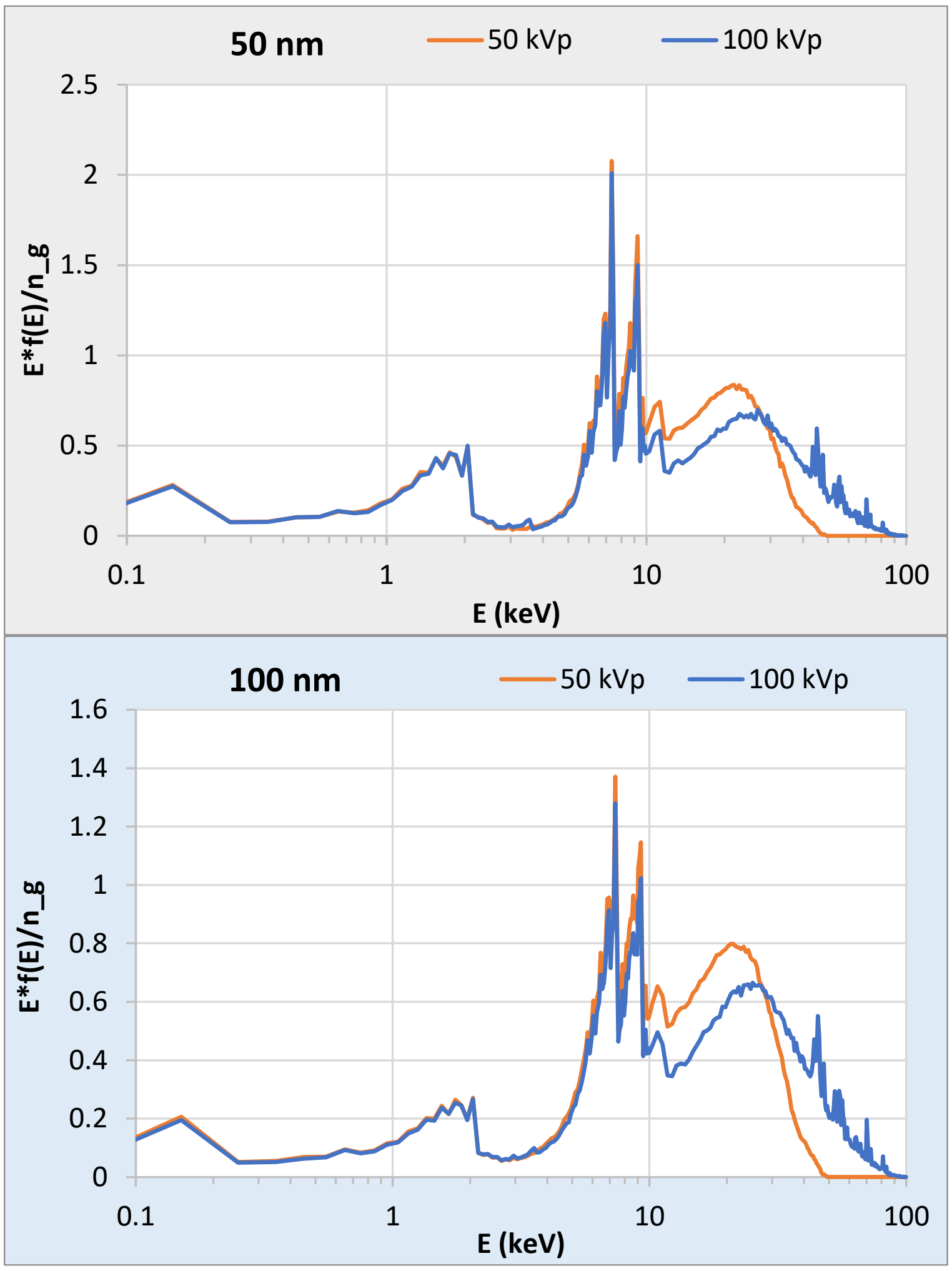

Figure 4: Comparison of the estimated energy flux of electrons emitted from a GNP that experienced a photon interaction for the two considered photon spectra and the 50 nm GNP (top) and the 100 nm GNP (bottom). (Data from participant NASIC.)

### *3.1.3 Consistency of electron spectra for the same GNP size and different X-ray spectra*

The photons of both energy spectra considered in the exercise have energies higher than 15 keV (Figure 2) and a mean free path in gold exceeding 3 µm (Figure 3). Therefore, their attenuation in the GNPs is negligible. In consequence, each atom in the GNP has the same probability of undergoing a photon interaction. Regarding the Auger electrons, the GNP may thus be considered as a uniform source. Furthermore, K-shell vacancies are mainly filled by a radiative transition which leave the excited ion with a hole in the L or higher shells, that is, in the same state as when a photoabsorption in these shells occurred. Therefore, it is expected that the L-, M- and N-shell Auger electron spectra for the same GNP diameter are about the same for the two radiation qualities. Comparing the electron for the same GNP size and the two X-ray spectra was therefore a first plausibility check.

Figure 4 shows results from a participant, which passed this plausibility test. (In fact, this participant was the only one whose results passed all consistency checks.)

The data shown in Figure 4 were normalized to the expected number of photon interactions in a GNP, $\bar{n}_g$, obtained from the following relation (Rabus *et al* 2021a):

$$\bar{n}_g = \frac{\pi}{6} {d_g}^3 \times \frac{1}{{d_b}^2 \pi / 4} \int \mu_{Au}(E) \times \Phi^{(p)}(E)\, dE \,/ \int \Phi^{(p)}(E)\, dE \qquad (1)$$

where $d_g$ and $d_b$ are the diameters of the GNP and the photon beam, respectively, $\mu_{Au}$ is the linear attenuation coefficient of gold, and $\Phi^{(p)}$ is the spectral fluence of primary photons. $\mu_{Au}$ was obtained from the data shown in Figure 3(a) by multiplying with the mass density of gold.

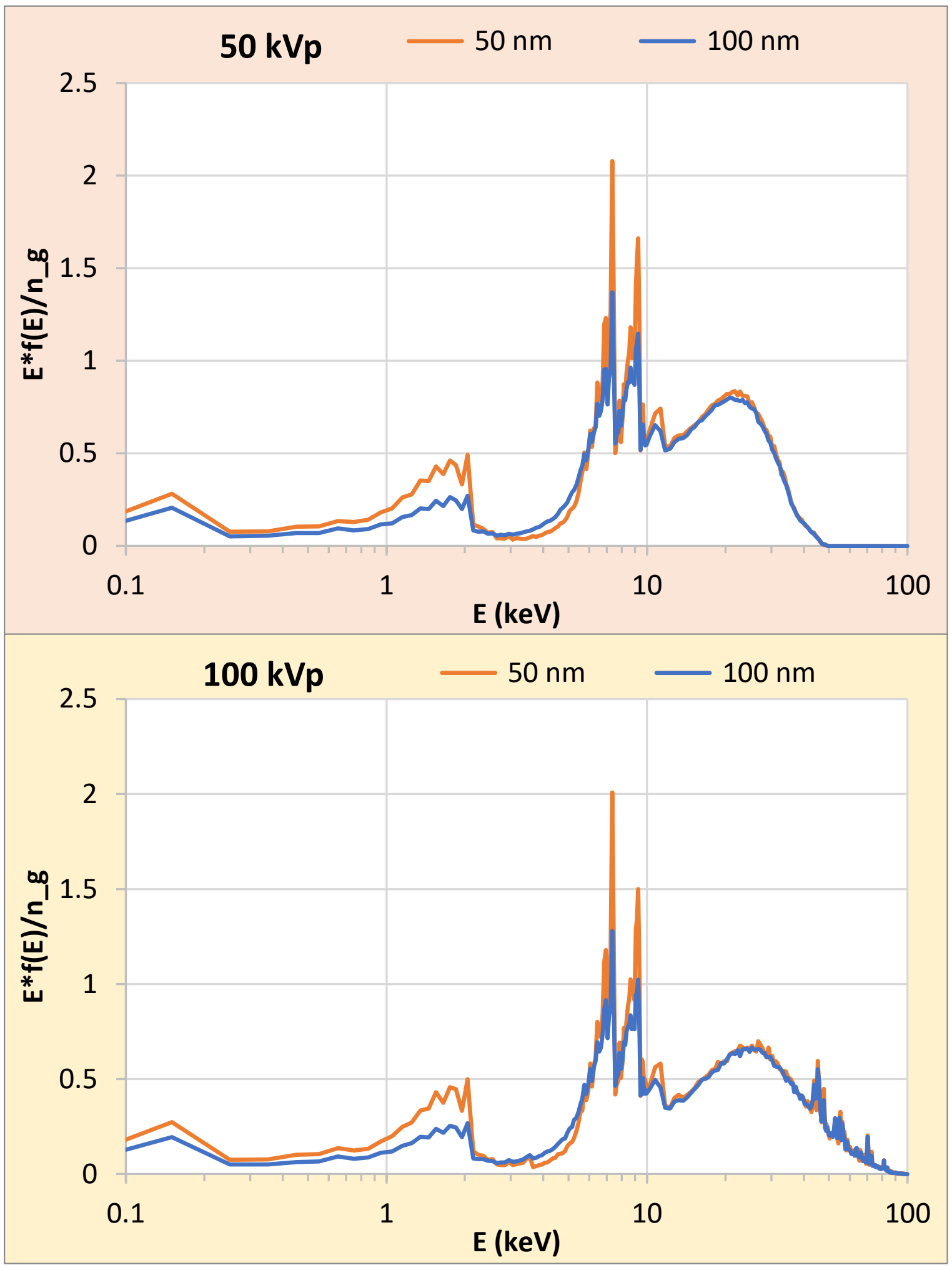

Figure 5: Comparison of the estimated energy flux of electrons emitted from a GNP that experienced a photon interaction for two considered GNP sizes and the 50 kVp photon spectrum (top) and 100 kVp spectrum (bottom). (Data from participant NASIC.)

Failing this plausibility test is an indication of a wrong implementation of the simulation geometry. Such a deviation from the exercise definition was identified and confirmed for the results of two participants (see Subsection 7.4). Results of a third participant also failed this test but were not confirmed and together with failure of further plausibility checks resulted in the exclusion of these results when the final reference values were established (Section 4).

### *3.1.4 Consistency of electron spectra for the same X-ray spectrum and different GNP size*

The electrons released in the GNP by photon interactions or ensuing de-excitation lose part of their energy before leaving the GNP by inelastic interactions. Inelastic electron interactions comprise inter alia impact ionization and bremsstrahlung emission. For metals, additional processes are plasmon and phonon generation, while for water electronic excitations as well as excitation of rotational, translational or vibrational modes of the molecular contribute (Salvat, 2015; Paretzke, 1987).

These energy losses change the energy spectrum of electrons leaving the GNP from that of electrons released by photon interactions. Since the energy losses are more important when the electron energy is smaller, it is therefore expected that when spectra of electron leaving the GNP are compared for the same X-ray spectrum and GNPs of different size, a reduced electron flux is expected for the larger GNP at low electron energies. On the contrary, changes at high electron energy should be negligible. This is seen in Figure 5, which shows the same data as Figure 4 but this time grouped according to the photon spectrum.

Failing this plausibility test is another indication of a wrong implementation of the simulation geometry and was found with the same results as the test described in Subsection 3.1.3.

## 3.2 Energy deposited in the scoring region in absence of the GNP

### *3.2.1 Physical background*

The scoring region (see Figure 1) is a sphere of radius 50 µm + GNP radius which is large enough to ensure longitudinal secondary electron equilibrium (i.e., in the direction of the primary photons). For the case of water only, the total energy deposited within all spherical shells, $\bar{E}_{w,d}^{(p)}$, should have a value comparable to the total energy released by photon interactions in the volume traversed by the primary beam, $E_{w,r}^{(p)}$. The two quantities are given by Eqs. (2) and (3)respectively.

$$\bar{E}_{w,d}^{(p)} = \int_{r_g}^{R} \bar{\varepsilon}_w(r)dr \tag{2}$$

$$E_{w,r}^{(p)} = \rho_w \times {r_b}^2\pi \times 2R \times \int E \times \frac{\mu_{en,w}(E)}{\rho} \times \Phi^{(p)}(E)dE \tag{3}$$

$r_g$ is the GNP radius and $\bar{\varepsilon}_w(r)$ is the average imparted energy per primary photon obtained in the simulations without the GNP. $\rho_w$ is the mass density of water, $r_b$ is the radius of the photon source, $R$ is the radius of the outer surface of the outermost spherical shell included in the scoring, $E$ is the photon energy, $\mu_{en,w}(E)/\rho$ is the mass energy absorption coefficient of water, and $\Phi^{(p)}$ is the spectral fluence (particles per area and energy interval) of photons in the region of interest (i.e., around the GNP position). The figure of merit for the first quantitative consistency check is therefore the ratio $C_w$ as defined in eq. (4)

$$C_w = \bar{E}_{w,d}^{(p)}/E_{w,r}^{(p)} \tag{4}$$

The denominator and numerator in eq. (4) depend on *R*., so this is also the case for the fraction. Figure 6 shows sample results for the variation of $C_w$ with parameter *R*, where it is evident that the curves converge asymptotically to a constant value close to unity. It should be noted, however, that neither of the curves shown in Figure 6 reaches this asymptotic value.

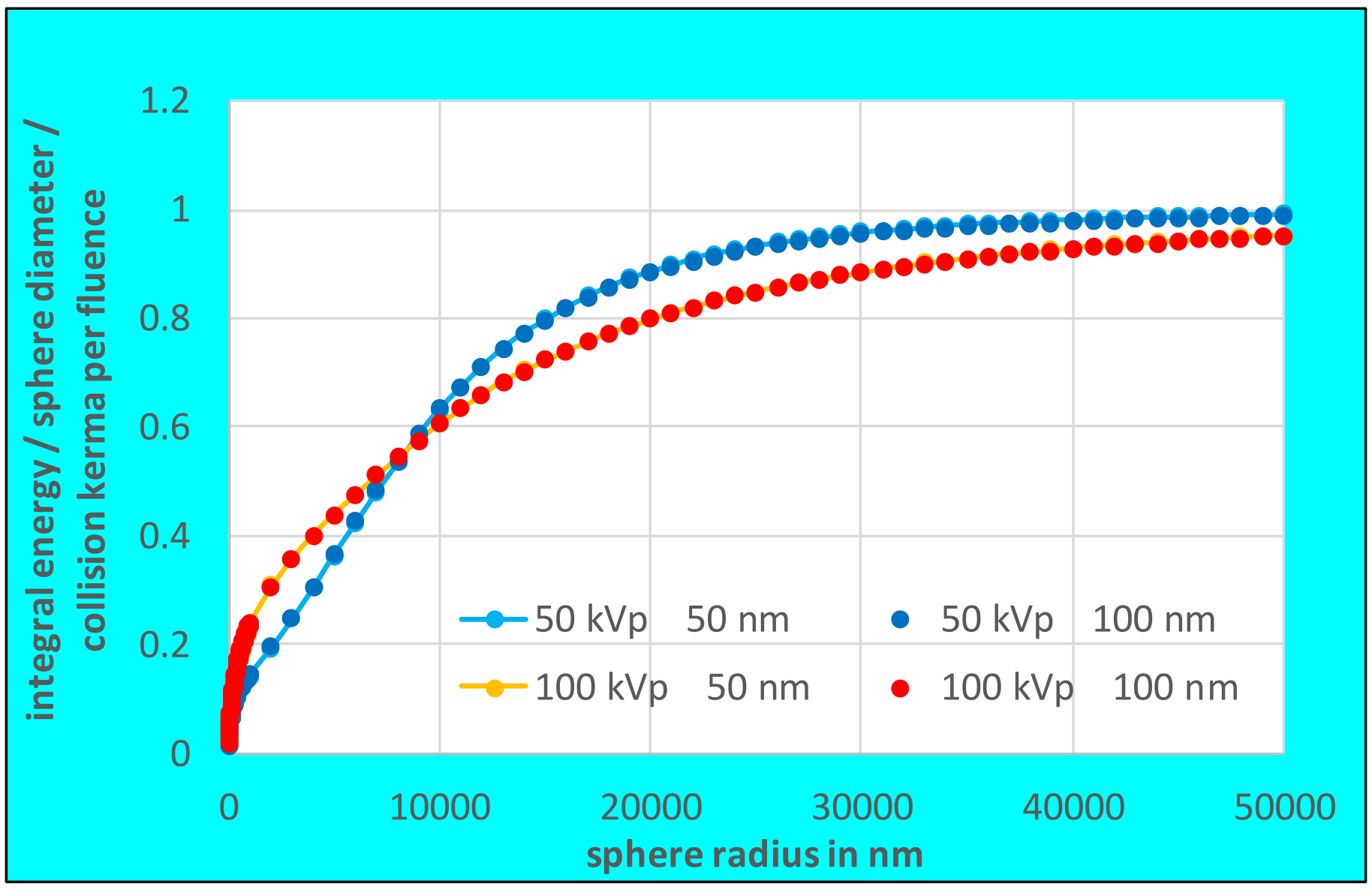

Figure 6: Ratio $C_W$ of the average dose within a water sphere (calculated from eq. (2)) and the collision kerma within the part of the sphere traversed by the primary photon beam (calculated from equation (3)). The data shown are based on the results from participant NASIC. For both photon spectra the curves for the two GNP sizes coincide.

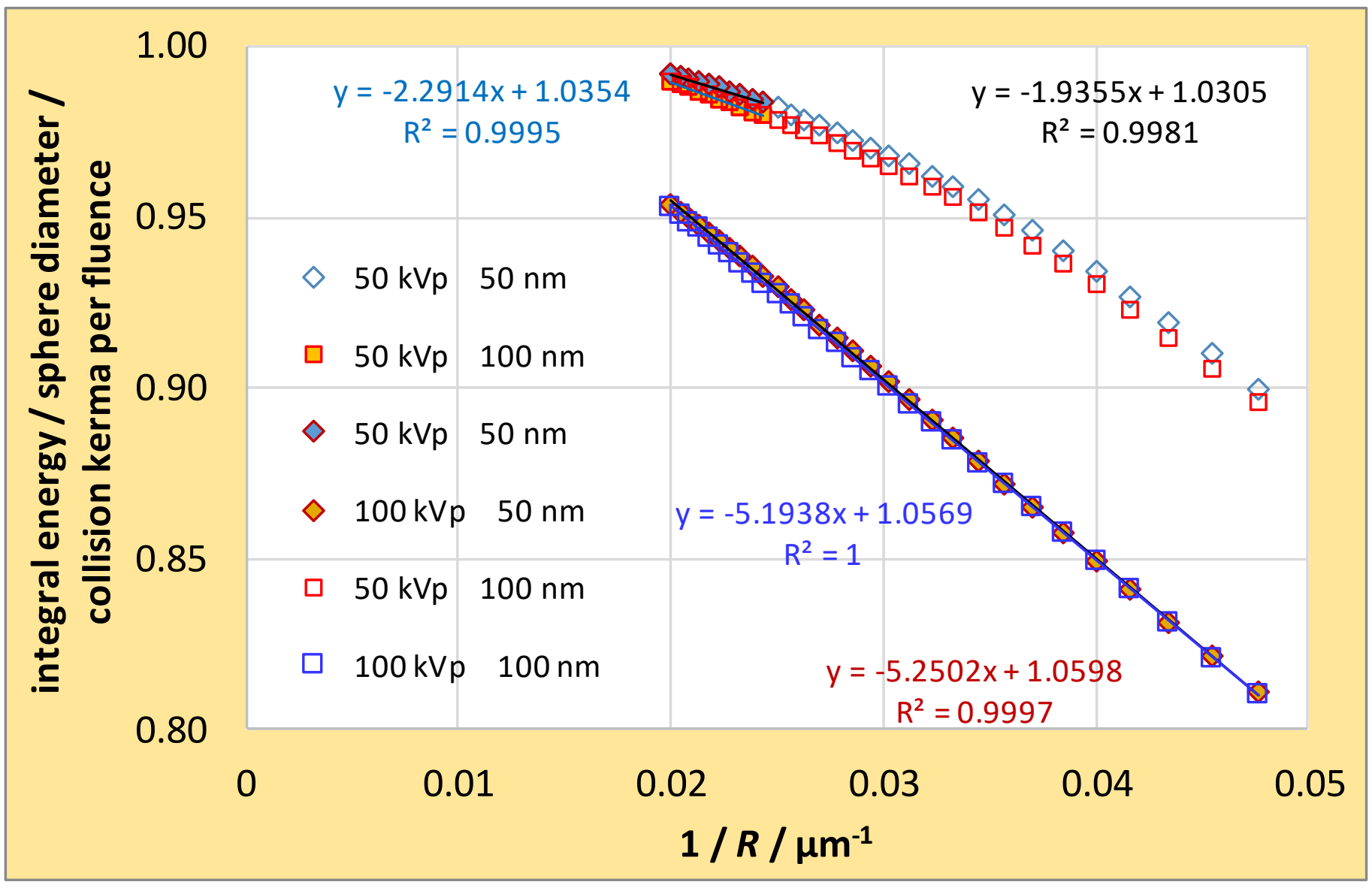

Figure 7: Illustration of extrapolation of the data shown in Figure 6 to the limit of $R$ going to infinity by linear regression as a function of $1/R$. The regression was performed taking into account only the filled symbols for each combination of radiation spectrum and GNP size.

In fact, the asymptotic value of $C_w$ is expected to be somewhat larger than unity, because there is also a contribution of energy deposition from electrons produced by interactions of photons that have been scattered out of the photon beam and their descendant photons. (That the volume of the GNP is not scored leads to a slight reduction, but as this volume is less that $10^{-9}$ of the total volume and less than $10^{-3}$ of the volume traversed by the primary beam, this is negligible.) To determine this asymptotic value, the variation of $C_w$ with $R$ at large values of $R$ was heuristically fitted assuming a $1/R$-dependence of the difference from the asymptotic value. In practice, this was done by doing a linear regression of $C_w$ as a function of $1/R$. (see Figure 7)

### *3.2.2 Outcomes for the initially reported data*

The results of this consistency check for the originally reported data are reported in Table 5 and Table 6, where cells highlighted in red indicate values that failed this consistency check. The reasons are discussed in Section 7.

Table 5: Asymptotic value of ratio $C_w$ (average dose within a sphere to the collision kerma within the part of the sphere traversed by the primary photon beam). The highlighted cells indicate values that fail the consistency check for being below unity or deviating too much from unity (Rabus *et al* 2021a).

| Participant ID | 50 kVp photon spectrum | | 100 kVp photon spectrum | |
|---|---|---|---|---|
| | 50 nm GNP | 100 nm GNP | 50 nm GNP | 100 nm GNP |
| G4/DNA#1 | 1.38 | 1.40 | 1.41 | 1.42 |
| G4/DNA#2 | - | - | - | - |
| G4/DNA#3 | 1.03 | 1.03 | 1.05 | 1.05 |
| MCNP6 | 1.08 | 1.02 | 1.04 | 1.04 |
| MDM | $6.90\times10^{7}$ | $2.04\times10^{7}$ | $6.67\times10^{7}$ | $1.92\times10^{7}$ |
| NASIC | 1.03 | 1.03 | 1.05 | 1.05 |
| PARTRAC | 0.85 | 0.84 | 0.85 | 0.84 |
| PENELOPE#1 | 0.75 | 0.73 | 0.81 | 0.79 |
| PENELOPE#2 | 0.71 | 0.73 | 0.81 | 0.81 |
| TOPAS | 1.03 | 1.03 | 1.05 | 1.05 |

Table 6: Regression coefficient for the extrapolation of the ratio $C_w$ defined in eq. (4) by linear fitting as a function of $1/R$. The highlighted cells indicate values where this linear regression does not seem to be a good representation of the data.

| Participant ID | 50 kVp photon spectrum | | 100 kVp photon spectrum | |
|---|---|---|---|---|
| | 50 nm GNP | 100 nm GNP | 50 nm GNP | 100 nm GNP |
| G4/DNA#1 | 0.9983 | 0.9993 | 0.9982 | 0.9997 |
| G4/DNA#2 | - | - | - | - |
| G4/DNA#3 | 0.9985 | 0.9997 | 0.9994 | 1.0000 |
| MCNP6 | 0.8480 | 0.9976 | 0.9985 | 0.9985 |
| MDM | 0.9966 | 0.9972 | 0.9994 | 0.9996 |
| NASIC | 0.9982 | 0.9995 | 0.9993 | 0.9998 |

| Participant ID | 50 kVp photon spectrum | | 100 kVp photon spectrum | |
|---|---|---|---|---|
| | 50 nm GNP | 100 nm GNP | 50 nm GNP | 100 nm GNP |
| PARTRAC | 0.9968 | 0.9848 | 0.9987 | 0.9978 |
| PENELOPE#1 | 0.9970 | 0.9880 | 0.9960 | 0.9962 |
| PENELOPE#2 | 0.9803 | 0.9864 | 0.9960 | 0.9962 |
| TOPAS | 0.9923 | 0.9924 | 0.9996 | 0.9996 |

## 3.3 Energy deposition from electrons leaving the GNP

### *3.3.1 Physical background*

If simulations are carried out using a photon beam of microscopic cross section, then the energy imparted in the scoring volumes is predominantly deposited by electron tracks produced within the region traversed by the primary photon beam. In the exercise, this region was a water cylinder of 100 µm height (diameter of the scoring region) with or without the presence of a GNP within.

In the presence of the GNP, there is additional energy deposition from electrons emerging from the GNP, which exceeds the energy deposition from electrons produced in the same volume in water (when the GNP is absent). The simulation results of deposited energy in the spherical shells around the GNP are basically the sum of aforementioned contributions and the energy depositions from electrons released by photon interaction in the rest of the volume traversed by the beam (except the sphere covered by the GNP or filled with water). Therefore, the difference of the simulation results for the cases with and without the GNP can be used as approximation for the extra dose contribution due to the GNP.

This assumption means that if $\bar{\varepsilon}_g(r)$ and $\bar{\varepsilon}_w(r)$ are the average imparted energies per primary photon in 10 nm spherical shells obtained in the simulations with the GNP and without the GNP, respectively, then the difference $\Delta\bar{\varepsilon}_g(r)$ given by

$$\Delta\bar{\varepsilon}_g(r) = \bar{\varepsilon}_g(r) - \bar{\varepsilon}_w(r) \quad (5)$$

is approximately giving the average extra energy imparted in this spherical shell by electrons that have been produced in the GNP.

Therefore, the quantity $\Delta\bar{\varepsilon}_g{}^*$ defined by eq. (6) is the additional mean imparted energy around a GNP in which photon interactions take place. $\bar{n}_g$ is the expected number of photon interactions (eq. (1)).

$$\Delta\bar{\varepsilon}_g{}^*(r) = \frac{\bar{\varepsilon}_g(r) - \bar{\varepsilon}_w(r)}{\bar{n}_g} \quad (6)$$

If there is no photon interaction in the GNP, this additional contribution is zero.

The radial integral, $\Delta\bar{E}_g{}^*$, calculated by eq. (7) is then the average total energy deposited in the water surrounding the GNP by electrons produced in the GNP when a photon interaction occurred in the GNP.

$$\Delta\bar{E}_g{}^* = \int \Delta\bar{\varepsilon}_g{}^*(r)\, dr \quad (7)$$

From energy conservation, this energy cannot be higher than the average energy transferred to electrons when a photon interaction takes place in the GNP.

The mean energy $\bar{E}_i$ transferred to electrons when a photon interaction in gold takes place can be estimated according to eq. (8).

$$\bar{E}_i = \frac{\int E\mu_{en,Au}(E)\Phi^{(p)}(E)e^{-\mu_w(E)d_g}dE}{\int \mu_{Au}(E)\Phi^{(p)}(E)e^{-\mu_w(E)d_g}dE} \quad (8)$$

where $E$ is the photon energy, $\mu_{en,Au}$ is the mass absorption coefficient of gold (Hubbell and Seltzer, 2004), $\Phi^{(p)}$ is the particle fluence of primary photons emitted from the X-ray source, $\mu_w$ and $\mu_{Au}$ are the photon mass attenuation coefficients of water and gold (Berger *et al.*, 2010) , and $d_g$ is the distance of the GNP from the photon source.

Most of the energy transferred to electrons by photon interactions in gold is transported outside the GNP by escaping electrons. These electrons then interact with the water medium, where the cross sections for radiative loss (bremsstrahlung) are much smaller than those for gold. Hence, escaping electrons will deposit a larger fraction of their energy by ionizing collisions, thus leading to a larger imparted energy than that for the case without gold. Therefore, $\bar{E}_i$ is an upper bound of the energy imparted to water around the GNP if a photon interaction takes place in the GNP.

The second figure of merit in the consistency checks is therefore the ratio $C_g$ as defined in eq. (9)

$$C_g = \Delta\bar{E}_g^{\,*}/\bar{E}_i \quad (9)$$

### *3.3.2 Outcomes for the initially reported data*

The results of this consistency check for the originally reported data are reported in Table 7. Cells highlighted in red indicate values that failed this consistency check. The causes for this failure are discussed in Section 7.

Table 7: Ratio $C_g$ of the average total excess energy deposited by electrons produced in a GNP that experienced a photon Interaction and the average energy transferred to electrons by a photon interaction in gold. The highlighted cells indicate values that fail the consistency check for being above unity or deviating too much from unity.

| Participant ID | 50 kVp photon spectrum | | 100 kVp photon spectrum | |
|---|---|---|---|---|
| | 50 nm GNP | 100 nm GNP | 50 nm GNP | 100 nm GNP |
| G4/DNA#1 | 1.19 | 1.11 | 1.19 | 1.13 |
| G4/DNA#2 | - | - | - | - |
| G4/DNA#3 | 1.27 | 1.01 | 1.27 | 1.03 |
| MCNP6 | 0.83 | 0.82 | 0.87 | *0.76*<br>0.83* |
| MDM | $5.41\times10^7$ | $1.50\times10^7$ | $5.27\times10^7$ | $1.51\times10^7$ |
| NASIC | 0.87 | 0.83 | 0.88 | 0.85 |
| PARTRAC | 1.18 | 0.84 | 1.21 | 0.91 |
| PENELOPE#1 | 0.54 | 0.51 | 0.39 | 0.37 |
| PENELOPE#2 | 0.53 | 0.49 | 0.38 | 0.36 |
| TOPAS | 0.65 | 0.71 | 0.65 | 0.72 |

* The number in italics refers to the originally delivered data that were superseded by revised results in the course of writing the publication (Li *et al.*, 2020a).

## 3.4 Energy transported out of the GNP by emitted electrons

### 3.4.1 Physical background

The reported spectra of ejected electrons from the GNP can also be used for consistency checks. The electrons leaving the GNP deposit their energy in the vicinity of the GNP and the extra energy imparted is the origin of the dose enhancement. Or, more quantitatively, the mean energy transported out of a GNP by emitted electrons, $\bar{T}^*$, is expected to be of about the same magnitude as $\Delta\bar{E}_g{}^*$ and $\bar{T}$ is given by

$$\bar{T}^* = \frac{1}{\bar{n}_g} \int T \times \Phi^{(e)}(T) dT \tag{10}$$

where $T$ is the kinetic energy of the electrons and $\Phi^{(e)}$ is their spectral flux (particles per energy interval). Both quantities can be used to estimate the fraction of the photon energy that is expended on energy deposits within the vicinity of the GNP, if a photon interaction occurs in the GNP.

On the other hand, the energy transported out of the GNP by electrons is only in part due to photon interactions with gold atoms. There is also a contribution originating from electrons that entered the GNP from outside and leave it again and from the secondary electrons produced by interactions of the entering electrons. Owing to these interactions, the energy transported out of the GNP by electrons that are not traceable to a photon interaction in the GNP is less than the energy transported into the GNP by entering electrons, such that there is a sink effect (Brivio *et al.*, 2017). Consequently, the net amount of energy transported out of the GNP by electrons is smaller than the energy from photon interactions occurring within in the GNP, such that $\bar{T}^*$ can be expected to be potentially less than $\bar{E}_i$.

Thus, the third consistency check is based on the ratio $C_e$ as defined in eq. (11)

$$C_e = \bar{T}^* / \bar{E}_i \tag{11}$$

### 3.4.2 Outcomes of the consistency check for the initially reported data

The results of this consistency check for the originally reported data are reported in Table 8. Cells highlighted in red indicate values that failed this consistency check. The reasons are discussed in Section 7.

Table 8: Ratio $C_e$ of the average total energy transported by electrons out of a GNP that experienced a photon Interaction and the average energy transferred to electrons by a photon interaction in gold. The highlighted cells indicate values that fail the consistency check for being above unity or deviating too much from unity.

| Participant ID | 50 kVp photon spectrum | | 100 kVp photon spectrum | |
|---|---|---|---|---|
| | 50 nm GNP | 100 nm GNP | 50 nm GNP | 100 nm GNP |
| G4/DNA#1 | 4.5 | 4.25 | 0.92 | 4.4 |
| G4/DNA#2 | 1.85E-03 | 2.16 | - | - |
| G4/DNA#3 | 3.14 | 2.46 | 6.40 | 5.20 |
| MCNP6 | *5.61*<br>3.77* | 3.56 | 4.10 | 3.63 |
| MDM | *$1.24\times10^{-8}$* | *$2.05\times10^{-5}$* | *$1.43\times10^{-8}$* | *$4.41\times10^{-8}$* |

| Participant ID | 50 kVp photon spectrum | | 100 kVp photon spectrum | |
|---|---|---|---|---|
| | 50 nm GNP | 100 nm GNP | 50 nm GNP | 100 nm GNP |
| | 6.14* | 10.42* | 0.73* | 14.98* |
| NASIC | 0.88 | 0.83 | 0.90 | 0.86 |
| PARTRAC | 0.88 | 0.83 | 0.90 | 0.87 |
| PENELOPE#1 | 1.20 | 0.51 | 0.44 | 0.43 |
| PENELOPE#2 | - | - | - | - |
| TOPAS | 0.66 | 0.72 | 0.68 | 0.74 |

* Numbers in italics refers to the originally delivered data that were subjected to a correction during the initial analysis that led to writing publication (Li *et al.*, 2020a).

## 3.5 Correction for deviation from the simulation geometry

In the consistency checks, the results are normalized to the mean number of photon interactions occurring in the GNP $\bar{n}_g$ calculated with Eq. (1). In the consistency checks, the results are normalized to the mean number of photon interactions occurring in the GNP $\bar{n}_g$ calculated with Eq. (1). Hence, if a different beam size was used in the simulations, this shows as differences between the electron spectra at high energies for the two GNP sizes with the same radiation quality. (For examples, refer to section 7.4).

If a variant geometry was consistently used in the simulations of a participant, this impacts electron spectra and excess energy deposited in the presence of the GNP in the same way. Thus, the values for $C_g$ and $C_e$ should be comparable for each combination of GNP size and photon spectrum but show larger variance among themselves for different of GNP size and photon spectrum (see Table 7 and Table 8 for examples).

For the purpose of the consistency checks, the variant source geometry can be accounted for by a factor $F_\Phi$ correcting the actual photon fluence used in simulations as given in eq. (12)

$$F_\Phi = \frac{A_{sim}}{A_{exc}} \tag{12}$$

Here, $A_{sim}$ is the actually used source area and $A_{exc}$ is the source area defined in the exercise.

However, in contrast to the case of a general normalization factor (where both, the simulations with and without GNP have to be scaled in the same way to get the correct values), a variant source geometry does only impact the simulation results for the case of a GNP present.

The reason is that in the case of a uniform medium, i.e., in the simulations without GNP present, the energy released on average by interactions of the primary photons is not depending on the source size or beam diameter. As long as the deviation of the source size used in the simulations departs only slightly from the source size defined in the exercise, the average energy deposition in the spherical shells for a fixed number of primary photons will thus change only negligibly. This also applies to the spherical shells close to the volume occupied by the GNP in the simulations with GNP present, because for the case of no GNP the energy deposited in these shells is the results of interactions of electrons that effectively are produced in a region with tens of micrometers longitudinal extension.

For the simulations with the GNP, however, the different fluence implies that the relative fraction of photons passing the GNP changes and thus the number of photons interacting in the GNP. Following the rationale explained in section 3.3, the difference between the energy deposition results difference $\Delta\bar{\varepsilon}_g(r)$ given by eq. (5) can be taken as the average extra energy imparted in a spherical shell by electrons that have been produced in the GNP. This contribution refers to the photon fluence used in the simulations and needs to be corrected to the fluence that was to be used in the exercise using the factor $F_\Phi$ given in eq. (12). Thus, the corrected deposited energy ratio is given by (Li *et al.*, 2020b; Rabus *et al.*, 2021b)

$$DER_{corr}(r) = 1 + F_\Phi \frac{\Delta\bar{\varepsilon}_g(r)}{\bar{\varepsilon}_w(r)} \quad . \tag{13}$$

Therefore, the corrected DER is related to the value from the simulations via the following relation

$$DER_{corr}(r) = 1 + F_\Phi(DER_{sim} - 1)\,, \tag{14}$$

and a similar relation holds for the correction of the energy imparted in each spherical shell:

$$\bar{\varepsilon}_{g,corr}(r) = \bar{\varepsilon}_w(r) + F_\Phi\big(\bar{\varepsilon}_g(r) - \bar{\varepsilon}_w(r)\big) \tag{15}$$

## 3.6 Summary of the issues identified with the initially reported results

The issues identified with the initially reported results, the causes identified (where applicable) and the final assessment of the results (after corrections where needed) are summarized in Table 9. For a more detailed description of the problems encountered, the reader is referred to Appendix 1 (Chapter 7).

Table 9: Summary of the problems encountered with the different results, corrections applied and assessment of the final results.Note: Case is used in this table to refer to a combination of GNP size and X-ray spectrum.

| Participant ID | Problems with initial results | Reason(s) identified | Final results |
|---|---|---|---|
| G4/DNA#1 | Energy deposition values with and without GNP appeared to high by a factor of about 1.3. | Energy deposition data was reported normalized to the spherical shell volume, calculated as the difference in volume of spheres with the inner and outer radius of the shells. In the calculation of the sphere volume the factor 4/3 was truncated to 1 owing to integer operation. | After correcting the reported values, the complete set of data of the participant passed all consistency checks. |
| | In three of four cases, the electron energy spectra appeared by a factor of about 4 to 5 too high. | The exercise organizer received the information that the spectra were reported as frequencies per bin but corrected only the values of one case. | |
| G4/DNA#2 | The participant reported only electron emission data for the 50 kVp X-ray spectrum which failed the consistency checks for being about a factor of 500 too low or a factor of 2.5 too high. | The reasons for the discrepancies could not be identified. The electron spectra appear physically implausible with missing M-shell Auger lines in one case and wrong energy positions of the L-shell Auger lines. (see 7.6.1) | The participant's data were excluded from the final dataset |

| Participant ID | Problems with initial results | Reason(s) identified | Final results |
|---|---|---|---|
| G4/DNA#3 | In contrast to energy deposition without GNP, which passed the consistency check, the values of energy deposition with GNP appeared enhanced by a factor of 1.4 for the 50 nm GNPs and by about 1.2 for the 100 nm GNP. | The consistency checks suggested that in the simulations a photon beam diameter equal to the GNP diameter was used (see 7.6.2).<br>This conjecture was not confirmed by the participant. | The participants' data failed the consistency checks and were excluded from the final dataset. |
| G4/DNA#3 | The electron spectra appeared to be too high by factors between 2.5 and 6.5 (depending on the combination of spectrum and GNP size). | The spectra were reported as frequencies per bin rather than frequency density. The consistency checks suggested that a photon beam diameter equal to the GNP diameter was used and that for the 50 kVp X-ray photon spectrum the requested GNP diameter was used for its radius as well as for the radius of the photon beam. This conjecture was not confirmed by the participant. | The participants' data failed the consistency checks and were excluded from the final dataset. |
| MCNP6 | In contrast to the energy deposition data, the electron spectra appeared by about a factor of 4 too high. | The spectra were reported as frequencies per bin rather than frequency density. In addition, only electron emitted within a small solid angle around the surface normal were scored. | After correcting the normalization to the bin width and applying a correction factor for the incorrect tally (empirically determined by the participant), the data pass all consistency checks. |

| Participant ID | Problems with initial results | Reason(s) identified | Final results |
|---|---|---|---|
| MDM | All reported data appeared by orders of magnitude too low. For the electron spectra, a tentative correction was applied, which led in three cases to values too high by an order of magnitude, and in one case to a value that was too low. | Instead of simulating photon transport, the participant used uniformly distributed electron sources in the GNP and water (representing the electrons produced by the interactions of the incident photon beam. In this approach, a photon fluence of 1/cm$^2$ was used instead of 1 photon per source area. The tentative correction for the electron spectra incorrectly assumed a different origin of the discrepancy. In addition, for one case the wrong data column had been picked from the results reported by the participant. | With the correct data column and the corrected normalization, the results pass all consistency checks except for the energy imparted in water in the absence of the GNP. The latter is due to the variant approach of using an electron source. |
| NASIC | None | Not applicable | All data passed the consistency checks |
| PARTRAC | In contrast to the electron spectra, which passed the consistency checks, the energy deposition without GNP appeared by a factor of 1.2 too low, whereas the energy deposition with the 50 nm GNP appeared by a factor of about 1.3 too high. | The participant did not provide further details of their simulations so that the reasons for the discrepancies could not be identified. | The participants' data failed the consistency checks and were excluded from the final dataset. |

| Participant ID | Problems with initial results | Reason(s) identified | Final results |
|---|---|---|---|
| PENELOPE#1 | The energy deposition values for the cases with and without the GNP appeared to be decreased by a factor of about 1.3. | There were two deviations from the exercise definition:<br>1. Instead of a circular beam of diameter 10 nm larger than the GNP, a rectangular beam having the requested size as the side length was used.<br>2. The cumulative distributions of the primary photon spectra were used instead of the photon spectra. | The participant repeated the results with a newer version of PENELOPE and the correct photon spectrum but the same geometry. After correcting the energy deposition with GNP and the electron spectra for the variant photon fluence, all results passed the consistency checks. |
| PENELOPE#2 | The energy deposition values for the cases with and without the GNP appeared to be decreased by a factor of about 1.3. (Electron spectra were not reported.) | The participant used the steering files of participant PENELOPE# and thus had the same two deviations from the exercise definition. | The participant withdrew their results. |
| TOPAS | In contrast to energy deposition without GNP, which passed the consistency check, both the energy deposition with GNP and the electron spectra appeared to be too low and showed an unplausible dependence on GNP size. | Instead of a beam of 10 nm larger diameter than the GNP, a 10 nm larger radius was used in the simulations. | After correcting the results for energy deposition with GNP and the electron spectra for the variant fluence, they passed the consistency checks |

# 4. Final exercise results

The exercise was conducted without a reference solution. In this section, the final results of the exercise are presented.

## 4.1 Summary of the outcomes of the consistency checks on the participant's final results

### *4.1.1 Energy deposited in the scoring region in absence of the GNP*

Table 10 shows the outcome of the consistency criterion for dose to water in the absence of the GNP. With the exception of participant PARTRAC, who did not revise the results, all reported results pass this test. For participant MDM the values are also below unity, but in this case, photons have not been tracked so that there is no contribution from scattered photons considered, so that a value slightly smaller than unity would be expected. For participant PENELOPE#1, the values for the 50 kVp spectrum are slightly smaller than unity, but this may be explained by potential differences between the cross sections used in the code and those in the XCOM database (Andreo *et al.* 2012).

Table 10: Asymptotic value of ratio $C_w$ (average dose within a sphere to the collision kerma within the part of the sphere traversed by the primary photon beam). The highlighted cells indicate values that fail the consistency check for being below unity or deviating too much from unity. (The values below unity of participant MDM are due to the fact that photons have not been tracked explicitly).

| Participant ID | 50 kVp photon spectrum | | 100 kVp photon spectrum | |
|---|---|---|---|---|
| | 50 nm GNP | 100 nm GNP | 50 nm GNP | 100 nm GNP |
| G4/DNA#1 | 1.04 | 1.05 | 1.06 | 1.06 |
| G4/DNA#2 | - | - | - | - |
| G4/DNA#3 | 1.03 | 1.03 | 1.05 | 1.05 |
| MCNP6 | 1.08 | 1.02 | 1.04 | 1.04 |
| MDM | 0.97 | 0.97 | 0.94 | 0.91 |
| NASIC | 1.03 | 1.03 | 1.05 | 1.05 |
| PARTRAC | 0.85 | 0.84 | 0.85 | 0.84 |
| PENELOPE#1 | 0.98 | 0.99 | 1.02 | 1.02 |
| PENELOPE#2 | - | - | - | - |
| TOPAS | 1.03 | 1.03 | 1.05 | 1.05 |

### *4.1.2 Energy deposition from electrons leaving the GNP*

Table 11 shows the results for the ratio between the total energy imparted in the water shells around the GNP to the total energy transported out of the GNP by emitted electrons. It can be seen that, with the exception of the two participants G4/DNA#3 and PARTRAC (for the 50 nm GNP), who did not revise their results, the final results of all other participants pass this consistency test.

Table 11: Ratio $C_g$ of the average total excess energy deposited by electrons produced in a GNP that experienced a photon Interaction and the average energy transferred to electrons by a photon interaction in gold. The highlighted cells indicate values that fail the consistency check for being above unity.

| Participant ID | 50 kVp photon spectrum | | 100 kVp photon spectrum | |
|---|---|---|---|---|
| | 50 nm GNP | 100 nm GNP | 50 nm GNP | 100 nm GNP |
| G4/DNA#1 | 0.89 | 0.84 | 0.89 | 0.85 |
| G4/DNA#2 | - | - | - | - |
| G4/DNA#3 | 1.27 | 1.01 | 1.27 | 1.03 |
| MCNP6 | 0.83 | 0.82 | 0.87 | 0.83 |
| MDM | 0.76 | 0.71 | 0.74 | 0.72 |
| NASIC | 0.87 | 0.83 | 0.88 | 0.85 |
| PARTRAC | 1.18 | 0.84 | 1.21 | 0.91 |
| PENELOPE#1 | 0.86 | 0.83 | 0.87 | 0.81 |
| PENELOPE#2 | - | - | - | - |
| TOPAS | 0.88 | 0.84 | 0.88 | 0.86 |

### *4.1.3 Energy transported out of the GNP by emitted electrons*

Table 12 shows the results for the ratio between the total energy transported out of a GNP in which a photon interacted to the average photon energy. Only the (unchanged) results of G4/DNA#3 as well as the revised results of participant G4/DNA#2 fail this consistency test.

Table 12: Ratio $C_e$ of the average total energy transported by electrons out of a GNP that experienced a photon Interaction and the average energy transferred to electrons by a photon interaction in gold. The highlighted cells indicate values that fail the consistency check for being above unity or deviating too much from unity.

| Participant ID | 50 kVp photon spectrum | | 100 kVp photon spectrum | |
|---|---|---|---|---|
| | 50 nm GNP | 100 nm GNP | 50 nm GNP | 100 nm GNP |
| G4/DNA#1 | 0.90 | 0.85 | 0.92 | 0.88 |
| G4/DNA#2 | $1.85\ 10^{-3}$ | 2.16 | - | - |
| G4/DNA#3 | 0.63 | 0.49 | 1.28 | 1.04 |
| MCNP6 | 0.83 | 0.79 | 0.90 | 0.80 |
| MDM | 0.86 | 0.87 | 0.95 | 0.91 |
| NASIC | 0.88 | 0.83 | 0.90 | 0.86 |
| PARTRAC | 0.88 | 0.83 | 0.90 | 0.87 |
| PENELOPE#1 | 0.85 | 0.81 | 0.88 | 0.84 |
| PENELOPE#2 | - | - | - | - |
| TOPAS | 0.90 | 0.85 | 0.92 | 0.88 |

## 4.2 Electron spectra

### *4.2.1 Spectral frequency density of emitted electrons per photon from the source*

In this Subsection, the final results of the exercise for the energy spectra of emitted electrons are presented. Including only results passing the consistency checks, a reference data set with an estimated uncertainty band is derived from these results. This reference dataset is made available as supplementary data to this report (see Appendix 2, Section 8.1). It is also included in modified excel workbooks as were used in the data analysis of the exercise, also available as supplements.

As mentioned earlier, different energy binning was used by the participants in the exercise. Most of the participants whose final results for the electron spectra passed the consistency checks used linear energy binning. The bin sizes used varied between 5 eV and 50 eV for the electron spectra produced by the 50 kVp spectrum and between 5 eV and 100 eV for the 100 kVp spectrum. Participant PARTRACK used logarithmic binning.

This non-harmonized reporting made a quantitative comparison difficult. Therefore, the spectra reported with a linear binning were rebinned in the electron energy range below 10 keV to the largest bin size of 100 eV used by participants. Since the statistics at higher energies was poor in all energy spectra, a larger bin size of 500 eV was used for rebinning in this energy range. To also include the results of participant PARTRACK reported with exponential energy binning, the cumulative distributions were linearly interpolated to obtain an approximate dataset with linear binning.

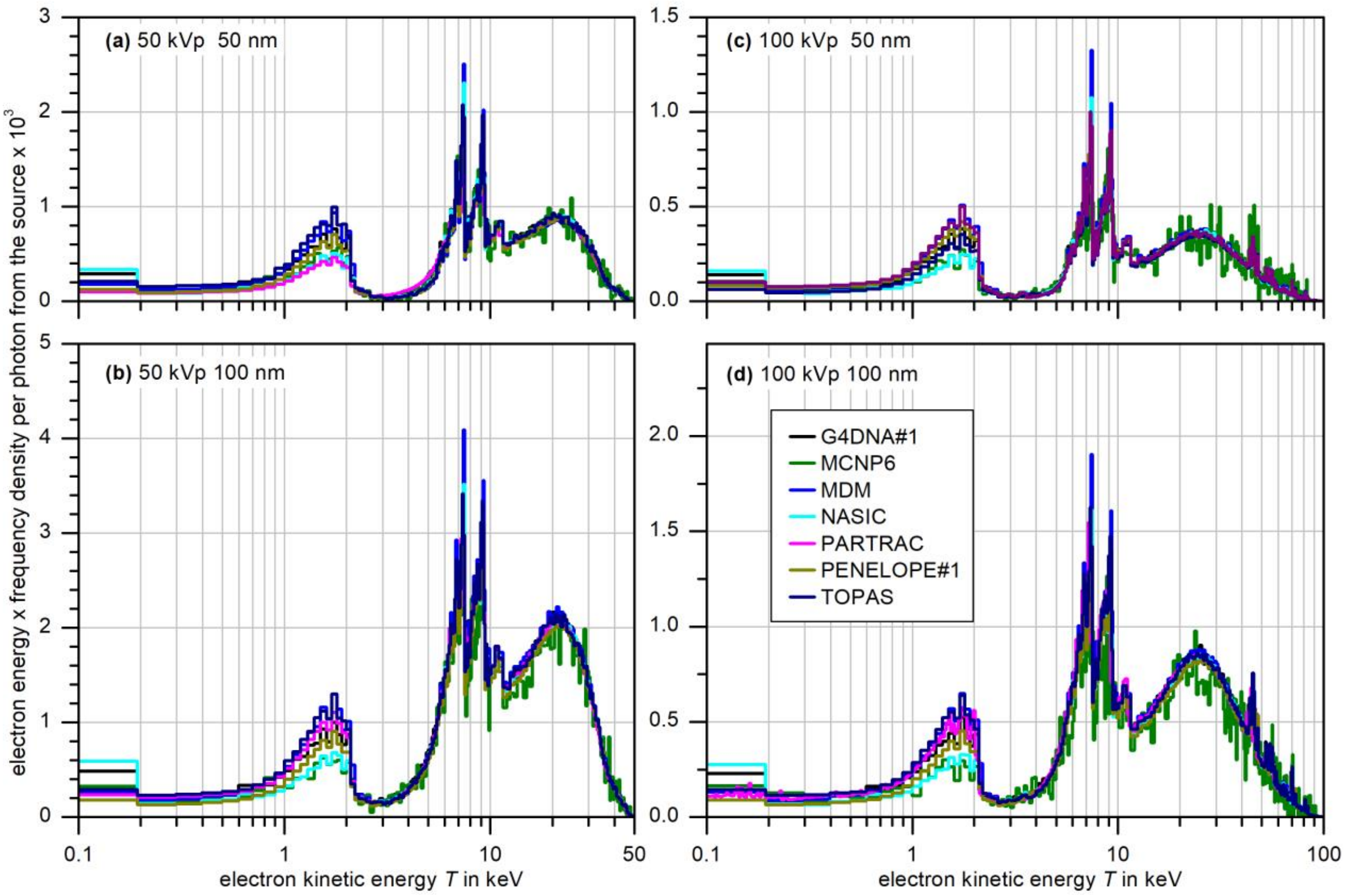

Figure 8: Energy spectra of electrons emitted per photon from the source for the two GNP sizes and X-ray spectra. These data correspond to the electron spectra reported in (Li *et al.*, 2020b), which have been rebinned and are presented in "microdosimetry style" (see text) so that the area under the curve is representative of the contributions from the respective energy range.

Figure 8 shows the final electron spectra for all combinations of GNP size and X-ray spectrum, reported as frequency density per photon emitted from the source area as defined in the exercise. They are plotted with a logarithmic x-axis and a linear y-axis showing the product of the frequency density and the energy. In this way of presentation, commonly used for microdosimetric spectra, the area under a part of the plotted spectrum is proportional to the contribution of that energy interval to the total integral under the curves.

In this presentation, it is obviously not possible to show the data at energies lower than 100 eV and to fully appreciate the differences between the different codes. Therefore, the final results of the participants at low energies (prior to re-binning) are shown in Figure 9. Evidently, the largest differences between participants occur in this energy range, highlighted by the logarithmic vertical scale to be used for showing all results. A quantitative assessment of a mean value and an uncertainty estimate is difficult for these non-homogenous data, so that they were not included in the reference data set derived from the final results that passed the consistency checks.

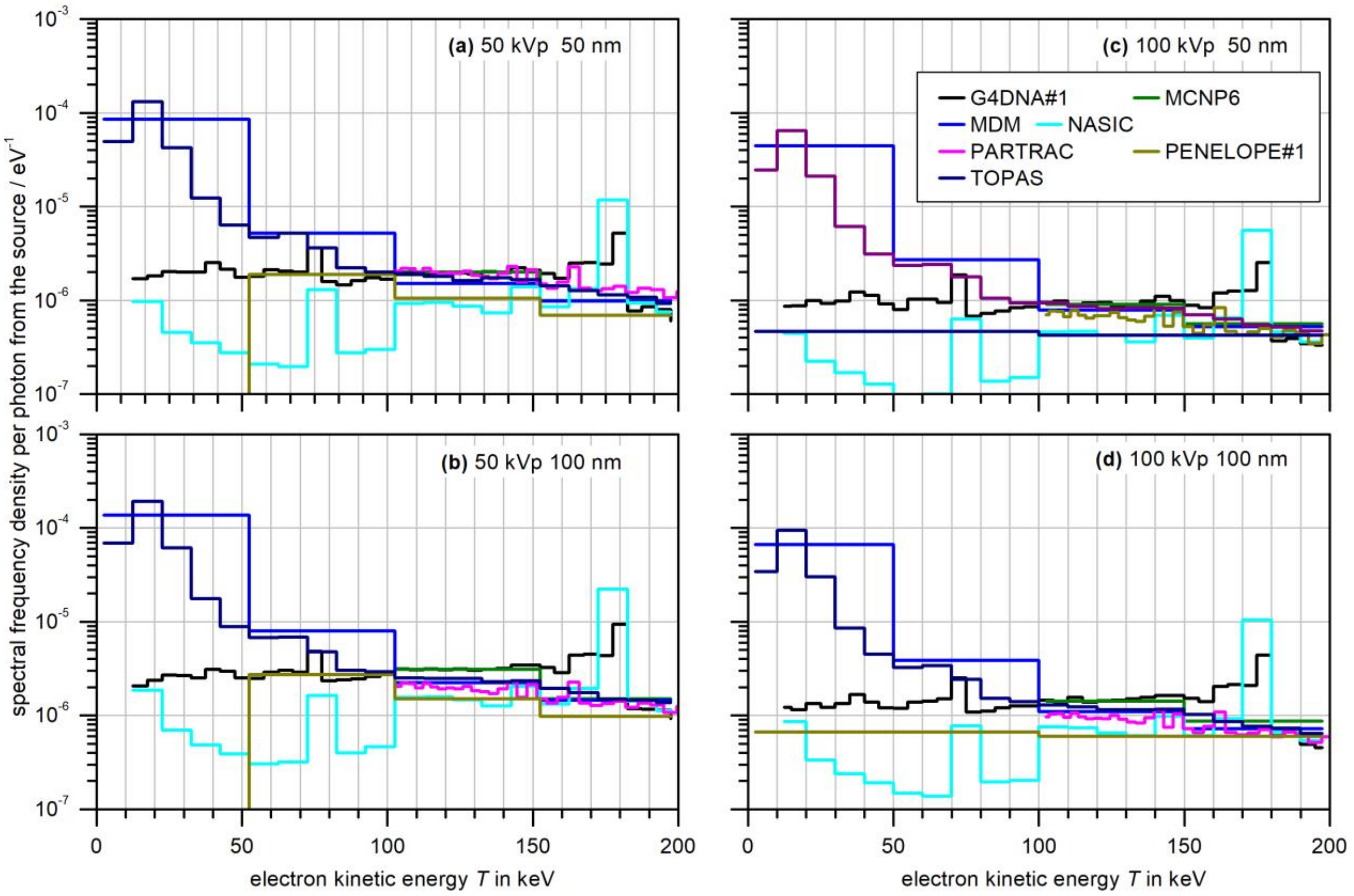

Figure 9: Comparison of the electron energy spectra in the energy range between 0 eV and 200 eV before rebinning. It should be noted that here the energy scale is linear and that a logarithmic vertical axis had to be used to accommodate the large differences between the results of different participants.

### *4.2.2 Reference dataset for the frequency density of electrons per photon from the source*

From the data shown in Figure 8, a reference dataset was constructed as the arithmetic mean of the participants' data, based on the assumption that the deviations between participants reflect unknown systematic uncertainty contributions dominating the overall uncertainty. The data of participant MCNP6 were excluded in the determination of the mean because this participant used a wrong tally of electrons emitted within a small solid angle about the surface normal. While the resulting bias could be corrected (Rabus et al. 2021b), a larger statistical uncertainty must be assumed owing to the inherently smaller count rate and is confirmed by the data shown in Figure 8.

An uncertainty band for the reference data set was established using the same approach as in Li et al. (2020a), namely assuming the different participants' values to originate from a log-normal distribution. The reference data set and the associated uncertainty band for 95 % uncertainty are presented in Figure 10.

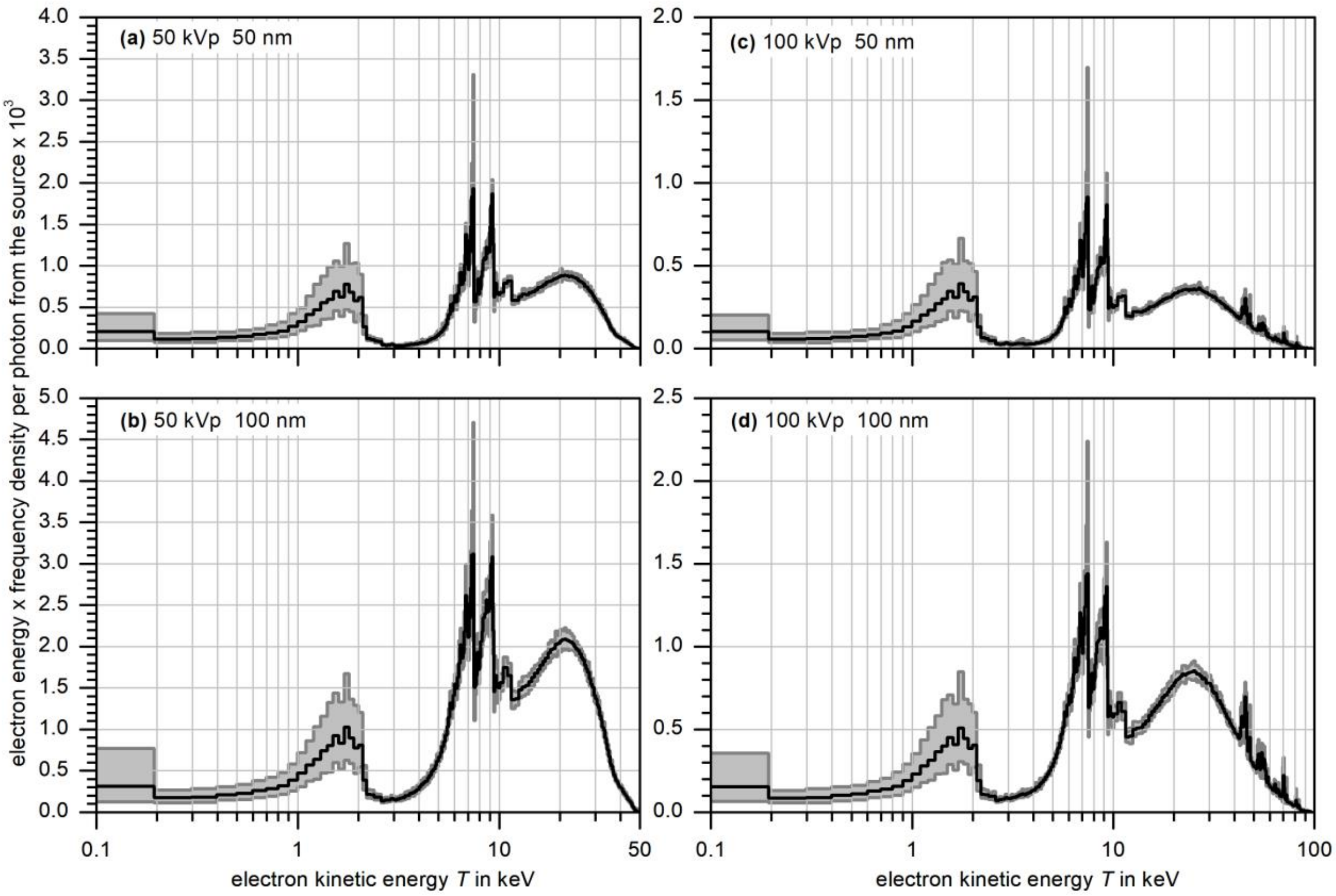

Figure 10: Reference values (black lines) and estimated uncertainty band for 95 % coverage (grey shaded areas) for the energy spectra of electrons emitted per photon from the source for the two GNP sizes and X-ray spectra. These data have been derived from the data shown in Figure 8 (see text).

### *4.2.3 Alternative presentations of the reference data for electron spectra*

Figure 11 shows the reference data for electron spectra for all combinations of GNP size and X-ray spectrum with a different normalization, namely to the mean number of interactions in the GNP (for a fluence of one photon per source area). These spectra represent the electron energy spectra when a photon interaction occurs in the GNP. This style of presentation removes the influence of the chosen value of photon fluence and thus highlights the dependence of the electron energy spectra on GNP size and X-ray spectrum: The spectrum at lower energies (below 3 keV) is mainly determined by the GNP size, whereas the spectrum at higher energies (above 10 keV) is mainly determined by the photon spectrum. The energy range between 3 keV and 10 keV comprises most of the gold L-shell Auger electrons and is influenced by both the GNP size and X-ray spectrum.

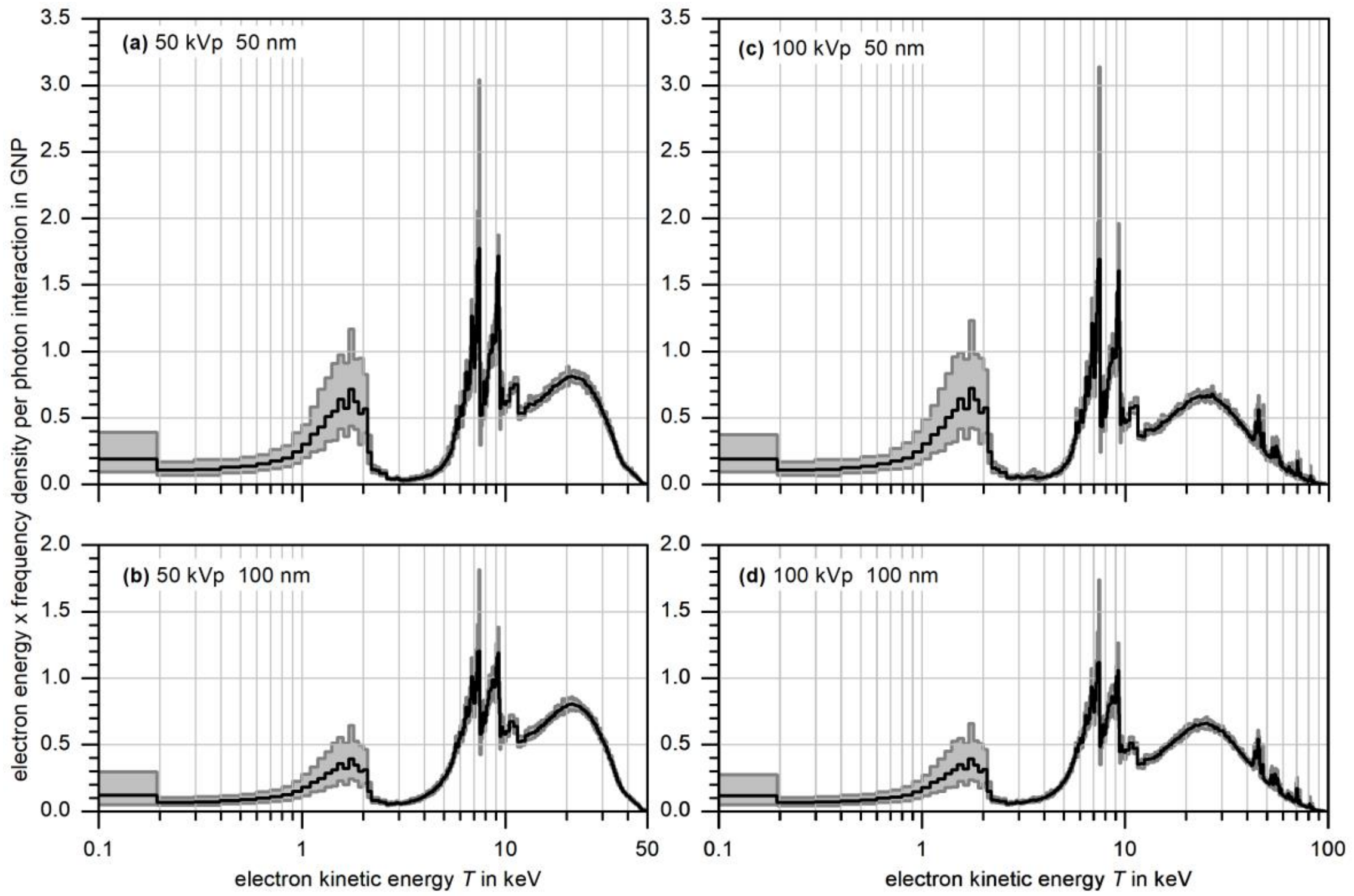

Figure 11: Same data as in Figure 10 but normalized to the mean number of photon interactions in the GNP.

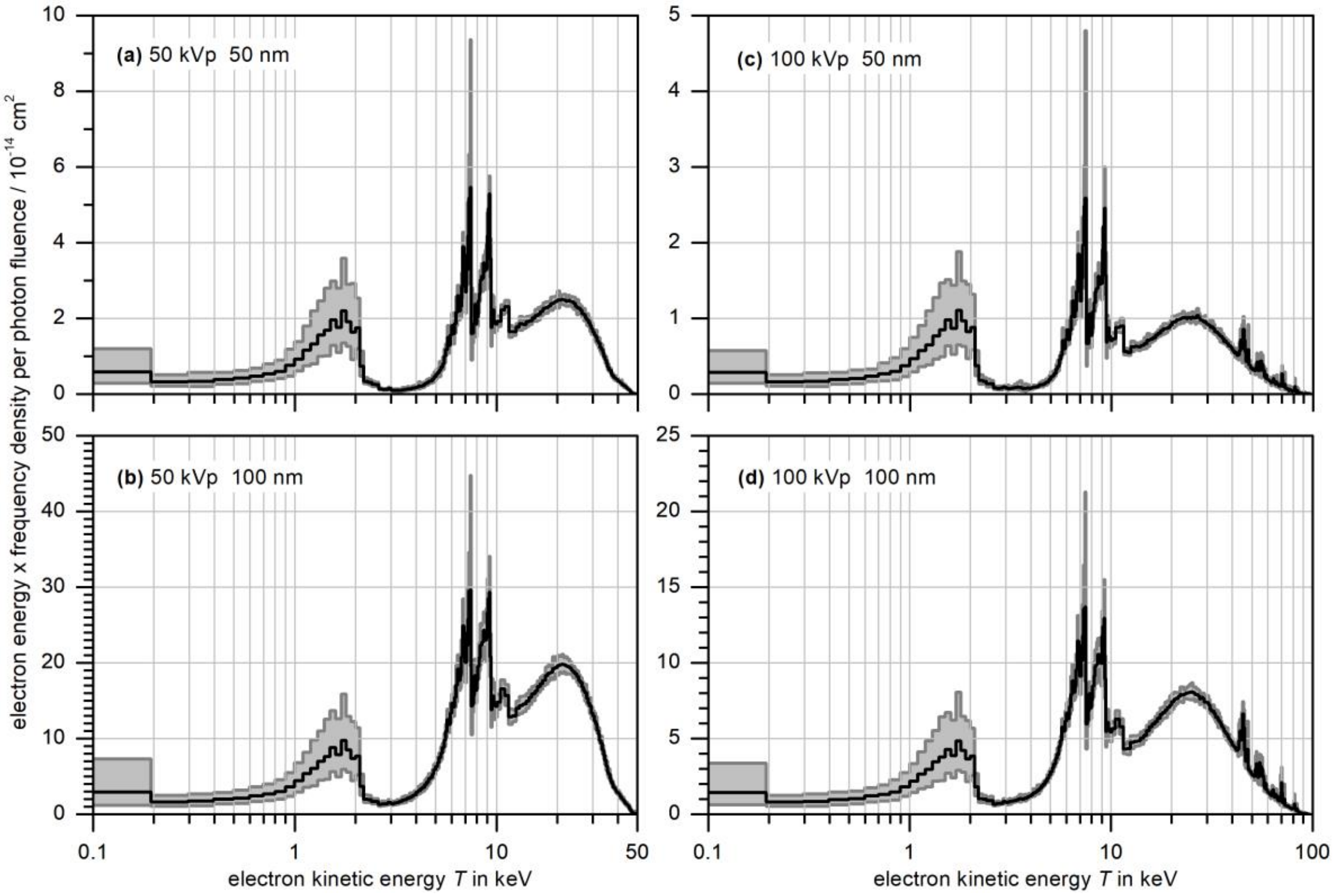

Figure 12: Same data as in Figure 10 but normalized to the fluence of primary photons.

Figure 12 shows the reference data for electron spectra normalized to the photon fluence. This style of presentation highlights the dependence of the electron spectra on GNP size and radiation quality.

Comparing the plots in the bottom row with those in the top row, the expected increase by a factor of 8 (corresponding to the increase in GNP volume) is seen at high electron energies (above 15 keV) for both radiation qualities. In contrast, the increase at lower energies is smaller and converging to the expected factor of 4 (increase of GNP surface area) with decreasing electron energy. Comparison of the plots on the left-hand side of Figure 12 with those on the right-hand side shows for both GNP sizes a decrease by a factor of about 2 for the electron spectra produced by the 100 kVp photon spectrum compared to those produced by the 50 kVp photon spectrum. This decrease reflects the decreasing interaction probability of the photons when their energy spectrum gets harder. It should be noted that the factor 2 agrees with the ratio of the two peak voltages only by coincidence. While a further decrease of the photon interaction probability and the resulting yield of electrons is expected for higher peak voltages, the reduction factor cannot be simply estimates from the peak voltage values.

## 4.3 Energy imparted in water shells surrounding the GNP

Figure 13 and Figure 14 show the energy imparted in water shells of thickness 10 nm and 1 µm, respectively, around the GNP or when the respective volume is filled with water (lower data in all panels). The data refer to the irradiation geometry used in the exercise, that is, a circular photon beam of diameter 10 nm larger than the GNP diameter.

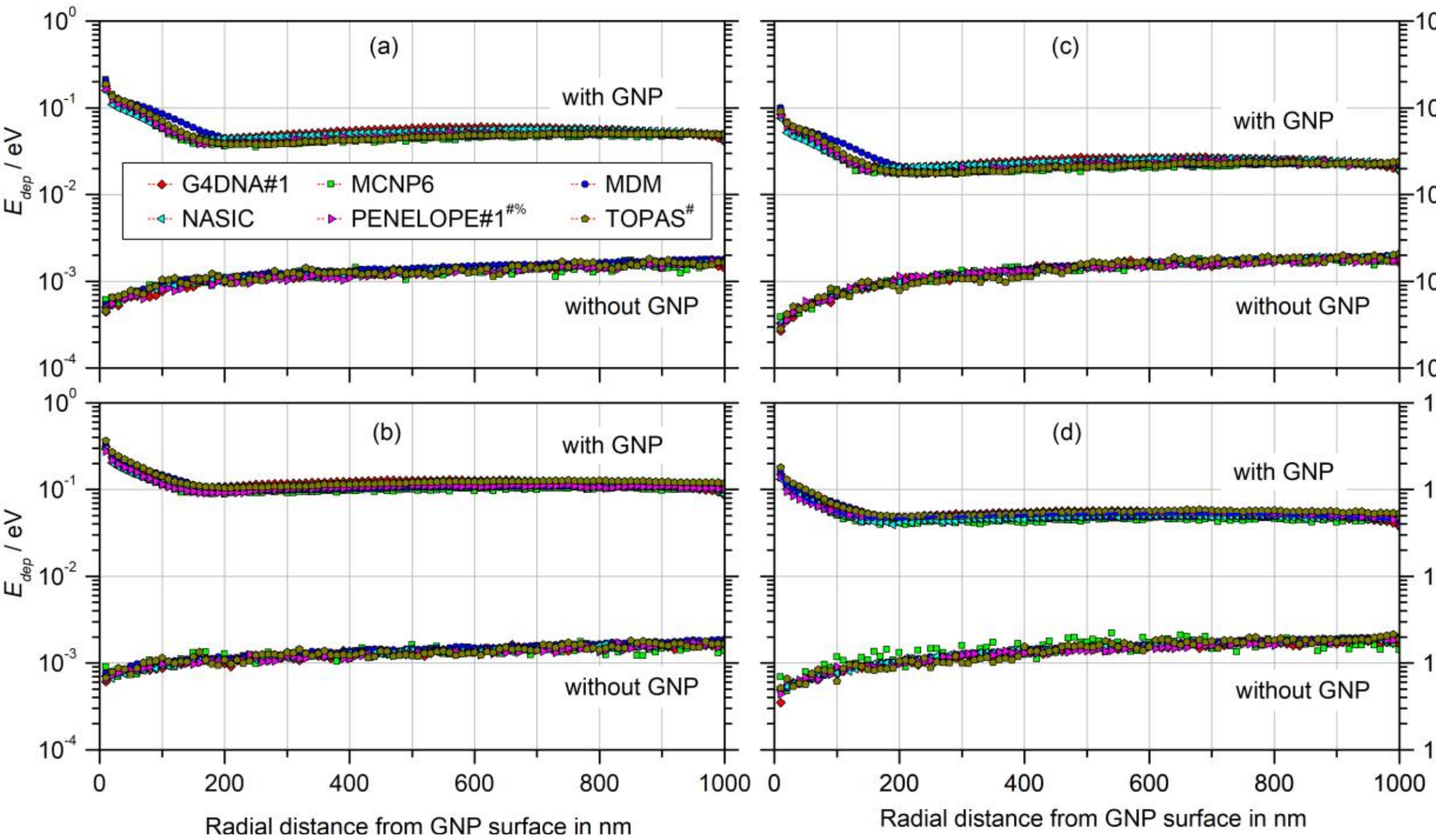

Figure 13: Energy imparted in 10 nm water shells around the volume occupied by the GNP (or filled with water) for distances from the surface of this volume up to 1000 nm. The data are normalized to the number of photons from the source and refer to the cases (a) 50 kVp, 50 nm GNP, (b) 50 kVp, 100 nm GNP, (c) 100 kVp, 50 nm GNP, (d) 100 nm GNP, 100 kVp.

The figures show only the data of participants whose results passed (in some cases after corrections) the two consistency checks for (a) the integral energy deposition with the total energy transported by electrons leaving the GNP and (b) the integral energy deposition in the absence of the GNP with the collision kerma. Corrections were applied to the data with GNP of participants PENELOPE#1 and

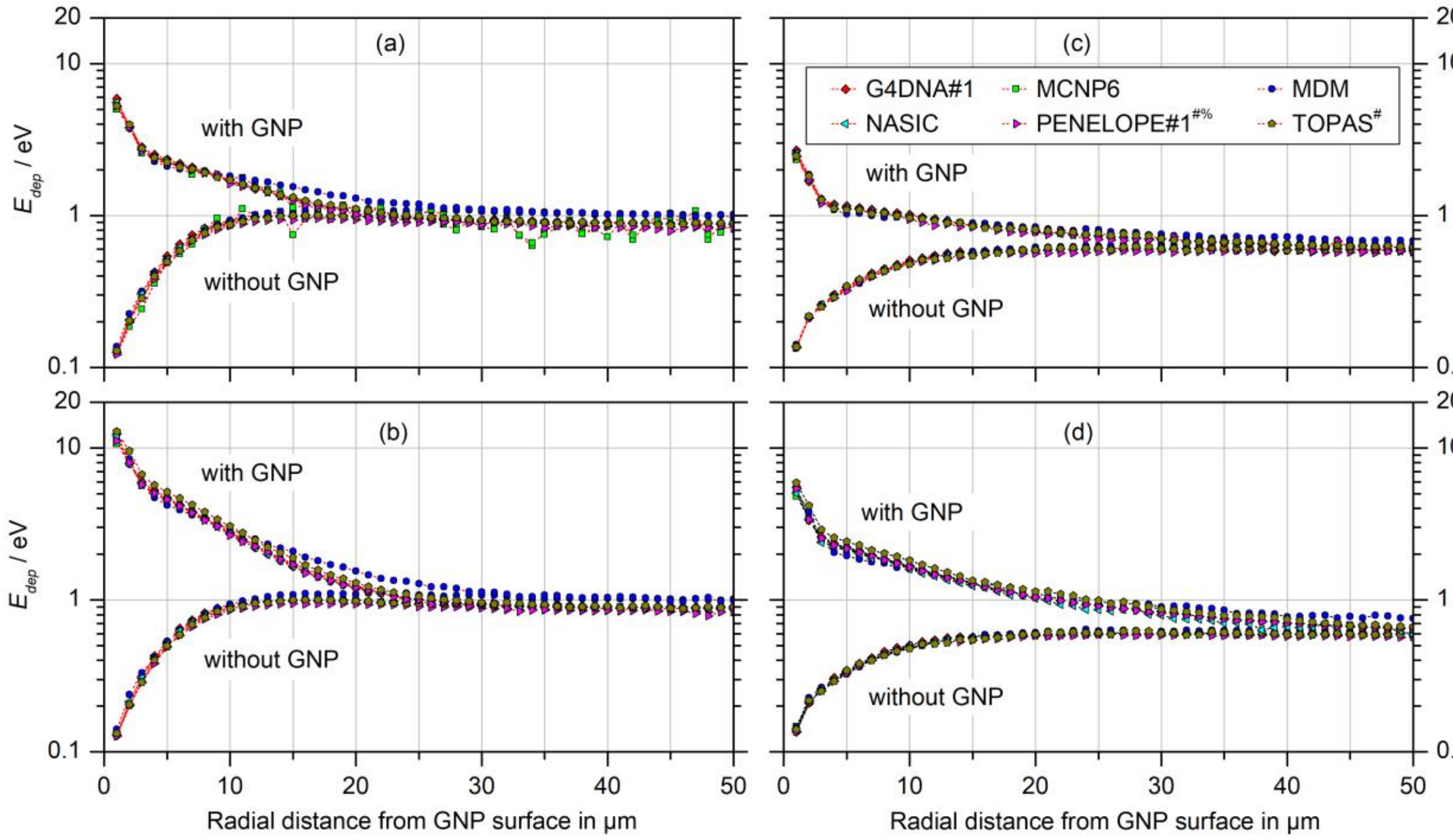

Figure 14: Energy deposition in 1 µm water shells around the volume occupied by the GNP (or filled with water). The data are normalized to the number of photons from the source and refer to the cases (a) 50 kVp, 50 nm GNP, (b) 50 kVp, 100 nm GNP, (c) 100 kVp, 50 nm GNP, (d) 100 nm GNP, 100 kVp.

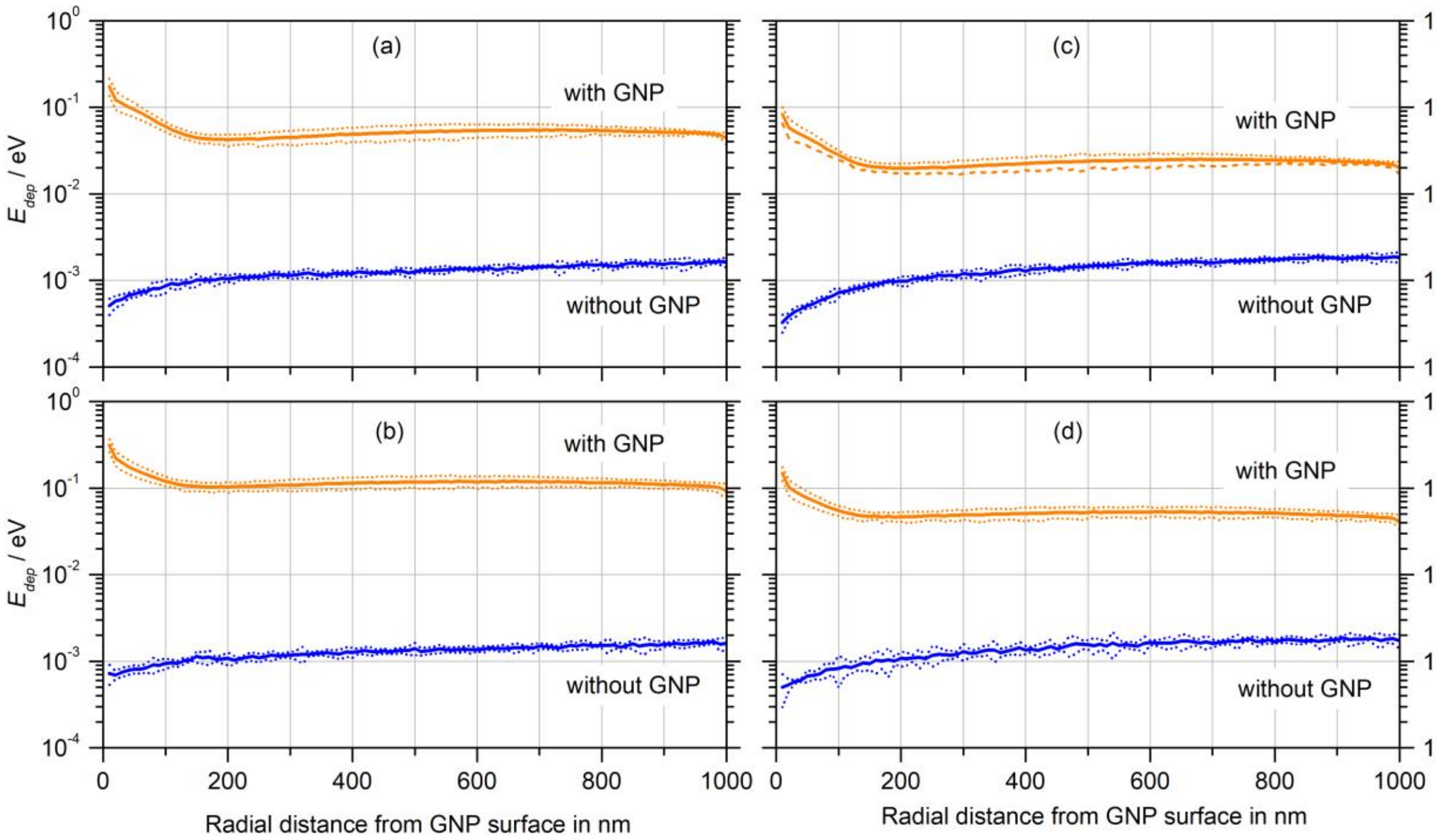

Figure 15: Estimates and uncertainty bands (see text) for the energy deposition in 10 nm water shells around the volume occupied by the GNP (or filled with water). The data are normalized to the number of photons from the source and refer to the cases (a) 50 kVp, 50 nm GNP, (b) 50 kVp, 100 nm GNP, (c) 100 kVp, 50 nm GNP, (d) 100 nm GNP, 100 kVp.

TOPAS according to Eq. (15) in Subsection 3.5 to account for a different beam shape and beam diameter, respectively, in the simulations of these participants.

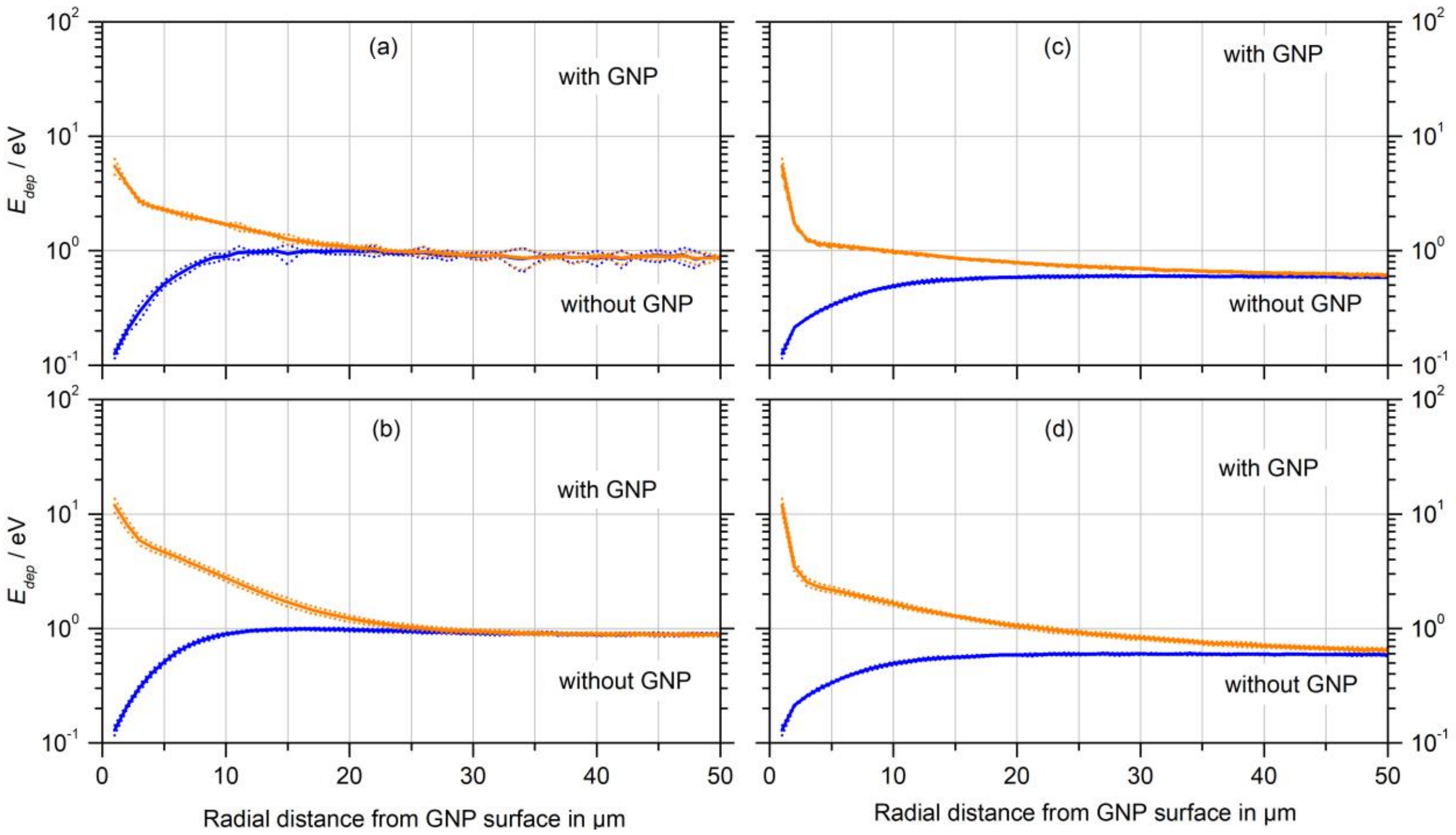

Figure 16: Estimates and uncertainty bands (see text) for the energy deposition in 1 µm water shells around the volume occupied by the GNP (or filled with water). The data are normalized to the number of photons from the source and refer to the cases (a) 50 kVp, 50 nm GNP, (b) 50 kVp, 100 nm GNP, (c) 100 kVp, 50 nm GNP, (d) 100 nm GNP, 100 kVp.

## 4.4 Caveats to the results

### *4.4.1 Deposited energy ratio versus dose enhancement*

Regarding the results for energy deposition around the GNP, it has to be emphasized that these are specific to the chosen narrow-beam geometry. The ratio of the energy deposited in the water shells with the GNP present and absent is not representative of the dose enhancement as the irradiation geometry does not ensure lateral secondary particle equilibrium (Rabus *et al* 2021b).

### *4.4.2 Cross-sections used in the simulations*

The results shown in this report are from simulations performed at the time the exercise was conducted, where PARTRAC and MDM were the only participating codes containing dedicated cross section datasets for low-energy electron transport in gold. In the meantime, discrete physics models for electron interactions in gold down to 10 eV have also been implemented within the Geant4-DNA low energy extension of Geant4 (Sakata *et al.*, 2016; Sakata *et al.*, 2018). Using these cross-section models in track structure simulation with Geant4 (and derivates such as NASIC and TOPAS) is expected to mainly have an impact on the low energy range of the emitted electron spectra and the resulting energy deposition in the immediate vicinity of the GNP.

### *4.4.3 Implementation of atomic relaxation*

Atomic relaxation is implemented in most Monte Carlo radiation transport codes using data from the evaluated atomic data library (EADL) (Perkins et al., 1991). To which extent the de-excitation cascade is followed depends on the set production cuts (Guatelli *et al* 2007). Some codes have additional restrictions. For instance, Penelope 2014 and 2018 stop the de-excitation cascades after all vacancies in the inner shells are filled (up to the N-shells). The EADL data refer to singly ionized atoms, while the atom is doubly ionized already after the first Auger transition. As shown by Pomplun (Pomplun *et al* 1987), relaxation of the electronic orbitals between the different steps of the de-excitation cascade leads to changes in transition energies and probabilities. This is not considered in the EADL. Furthermore, the data of high-energetic Auger transitions are mostly based on theoretical predictions. Recent experimental work found the gold and mercury L3 Auger energies in the EADL to be off by up to 100 eV (Rabus *et al* 2023).

### *4.4.4 Effects not considered in the simulation codes.*

GNPs can also emit low-energy electrons when surface plasmons , i.e., collective oscillations of free electrons on the surface of nanoparticles decay (Kuncic and Lacombe, 2018) . This effect is generally not considered in radiation transport codes.

For smaller GNPs, the emission of multiple low-energy Auger electrons may result in a highly charged and reactive nanoparticle, which may neutralize by extracting electrons from neighboring water and other biomolecules (Stumpf *et al.*, 2013), thereby amplifying radiation damage on neighboring molecules beyond that expected from emitted electrons alone (Li, 2006).

Finally, the simulations performed in the exercise were only considering energy deposition in water around GNPs. The biological effects of GNPs in irradiated cells are also due to chemical processes such as radical species from water radiolysis, which is enhanced in the vicinity of GNPs (Rudek *et al* 2019, Poignant *et al* 2020, 2021). Similarly, sub-excitation electrons can also undergo dissociative electron attachment to surrounding water and other biomolecules, causing further radiation-induced damage (Boudaiffa *et al.*, 2000).

# 5. Conclusions and perspectives

## 5.1 Conclusions

In this exercise, a simple simulation task of a single GNP with diameter of either 50 nm or 100 nm irradiated in liquid water by photons from an X-ray tube with either 50 kV or 100 kV peak voltage was solved applying seven renown Monte Carlo radiation transport simulation codes. Energy spectra of emitted electrons and the enhancement of deposited energy around the nanoparticle resulting from the interaction of the X-rays with the GNP were to be determined. Comparison of the results of different participants showed that even for the simple simulation setup, significant variations were obtained in the reported data, which in many cases could be traced back to deviations of the actual simulations from the requested geometry or other simulation settings by mistake. After applying corresponding corrections to data where the deviations could be explained and excluding results that still failed the consistency checks, large variations remain in the low-energy part of the energy spectra of electrons passing the GNP surface, especially in the energy range below 200 eV. (These electrons have ranges of few to several few tens of nanometres and therefore only make a difference in the immediate vicinity of the GNP surface.) These differences reflect the impact of different physical models and cross sections used in the codes as well as, of course, different cut-off energies and other physical parameters that have an impact on the simulation results. From the results that passed the consistency checks, datasets have been derived that reflect the spread of results to be expected in simulations solving the exercise tasks and may serve as a reference for newcomers in the field to test their capability of simulating such types of problems.

The simple irradiation geometry considered in the exercise is far from any realistic clinical irradiation scenario. Therefore, the results of this exercise should not be construed as being representative of what one would obtain with realistic irradiation setups. For instance, the setup shown in Figure 1 implies an irradiation with lateral particle disequilibrium. However, an approximate procedure was developed in the frame of the exercise to correct for this deficiency and to estimate the dose enhancement factor under condition of secondary particle equilibrium from the biased results obtained in the simulations (Rabus *et al.*, 2019, 2021b).

## 5.2 Perspectives for further exercise

The single GNP exercise showed a range of potential pitfalls and highlighted the need for quality assurance of MC simulations, for which several methods have been developed. With the experience gained in the single GNP exercise, a further exercise on multiple GNPs distributed in a cell can be considered, which would be more relevant for dosimetry in GNP-assisted radiotherapy and would probably result even in greater discrepancies between different codes and/or users. Besides the codes used in this exercise, other renowned MC codes, such as FLUKA (Böhlen *et al.*, 2014), EGSnrc (Kawrakow *et al.*, 2018) and PHITS (Sato *et al.*, 2018), may be invited to participate. It will also be of interest to study other quantities of interest such as clustering of molecular damage.

Such a new exercise can profit from the experience gained in this exercise regarding, inter alia, time frame and exercise management. The simulation setup should be simple and easy to implement in commonly used MC codes. The description of the exercise should be developed by a team rather than a single person to ensure that the description is clearly and understandably written. Several members of the coordinating consortium should first implement the exercise setup in their MC codes such as to identify potential ambiguities. Finally, clear rules for participation in the exercise should be established including a deadline for delivery of the results and provisions for copyright and authorship issues. Quantities to be reported must be clearly defined and should include the primary output quantities from the MC codes, from which other secondary quantities can be derived. The units to be used for the reported quantities must be very clearly stated to prevent confusion from participants using different units. For an efficient analysis of the delivered data, a reporting template should be provided for participants to fill their results. It is also advisable to define a set of criteria for checking the delivered results.

# 7. Appendix: Issues identified with the reported results

## 7.1 Overall normalization issues

### *7.1.1 Participant G4/DNA#1*

The energy deposition data reported by participant G4/DNA#1 failed the consistency checks for all cases of GNP size and photon spectrum, where the values seen in Table 5 and Table 7 appear intrinsically in agreement with each other and only enhanced by a factor of about 4/3. In fact, this participant reported simulation data for DER determination as energy deposited per volume of the radial shells (which is related to absorbed dose via the mass density of water). When the participant reviewed the script translating the simulation results, which were deposited energy per radial shell, it turned out that the factor 4/3 appearing in the radial shell volume was implemented as a ratio of integers so that integer truncation occurred (Rabus *et al.*, 2020). Thus, the results for energy per volume were by a factor of 4/3 too high, which accounts for the reported results failing to pass the consistency criteria in both Table 5 and Table 7. The revised data of this participant were the deposited energy per radial shell. Aforementioned normalization error has no impact on the DER value (for the narrow-beam geometry used in the exercise). It does only impact attempts to derive the DER for an extended photon field from the data obtained in the exercise (Rabus *et al.*, 2019).

### *7.1.2 Participant MDM*

For participant MDM, all originally reported data (energy deposition and electron spectra) failed the consistency tests where the discrepancy is by orders of magnitude. This suggested an overall normalization issue which could be tracked down to the particular simulation approach where photons were not tracked explicitly but rather a continuous electron source (based on photon interaction data) was used. In this approach, the fluence of photons was implicitly assumed to be $1/cm^2$ rather than 1 per photon source area as given in the exercise.

Rescaling the original data to the correct photon fluence made them pass the consistency checks with the exception of the electron spectra from a 100 nm GNP irradiated by 50 kVp photons. In this exceptional case it turned out that, in addition to the normalization issue, a wrong data column had been picked from the simulation output files in the initial data reporting. When the correct data column was used and corrected for the normalization problem, the data for this combination of GNP size and X-ray spectrum also passed the plausibility tests.

## 7.2 Energy bin normalization of electron spectra

As was mentioned earlier, electron spectra were reported by the participants using different bin size where only some participant reported spectral frequency (frequency per eV).

As seen in Table 5, the electron spectra of participant G4/DNA#1 fail the consistency tests in three of four cases. This is further illustrated in Figure 17 which shows (in the bottom left panel) that the electron spectra produced in the 50 nm GNP by the two X-ray spectra are qualitatively agreeing at energies below 10 keV in a similar degree as was seen in Figure 4. However, the absolute scale is by a factor around 5 different. Comparison with Figure 5 therefore suggests that the data for the 50 kVp spectrum and both GNP sizes as well as for the 100 nm GNP and the 100 kVp spectrum both have not been normalized to the energy bin width of 5 eV (Li *et al.*, 2020b). This was not a problem with the originally reported results, which were reported as frequency per photon per energy bin, but rather a (confirmed) processing error in preparing the figures for (Li *et al.*, 2020a).

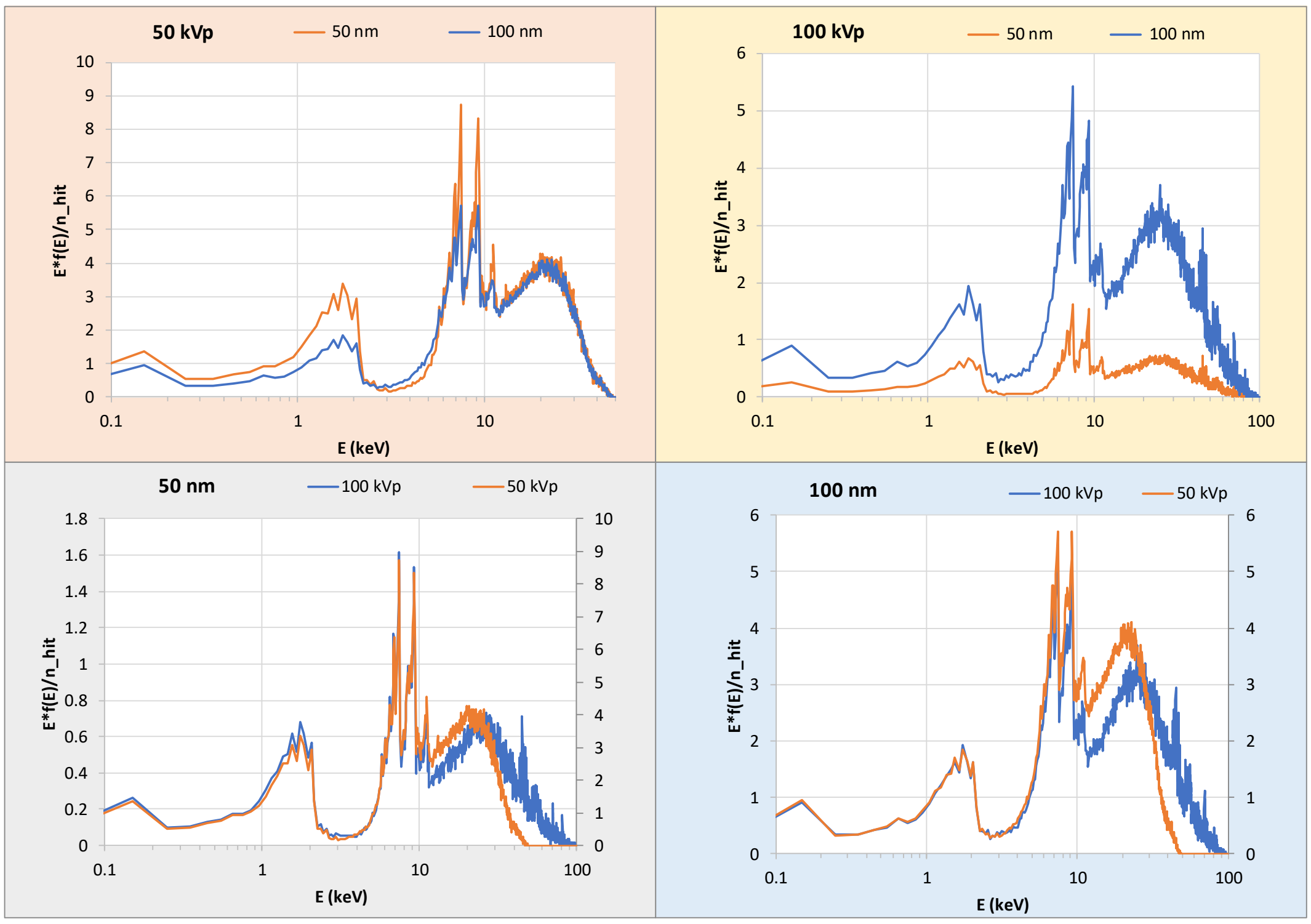

Figure 17: Electron energy flux per photon interaction in a GNP derived from the data shown in (Li *et al.*, 2020a) attributed to participant G4/DNA#1.

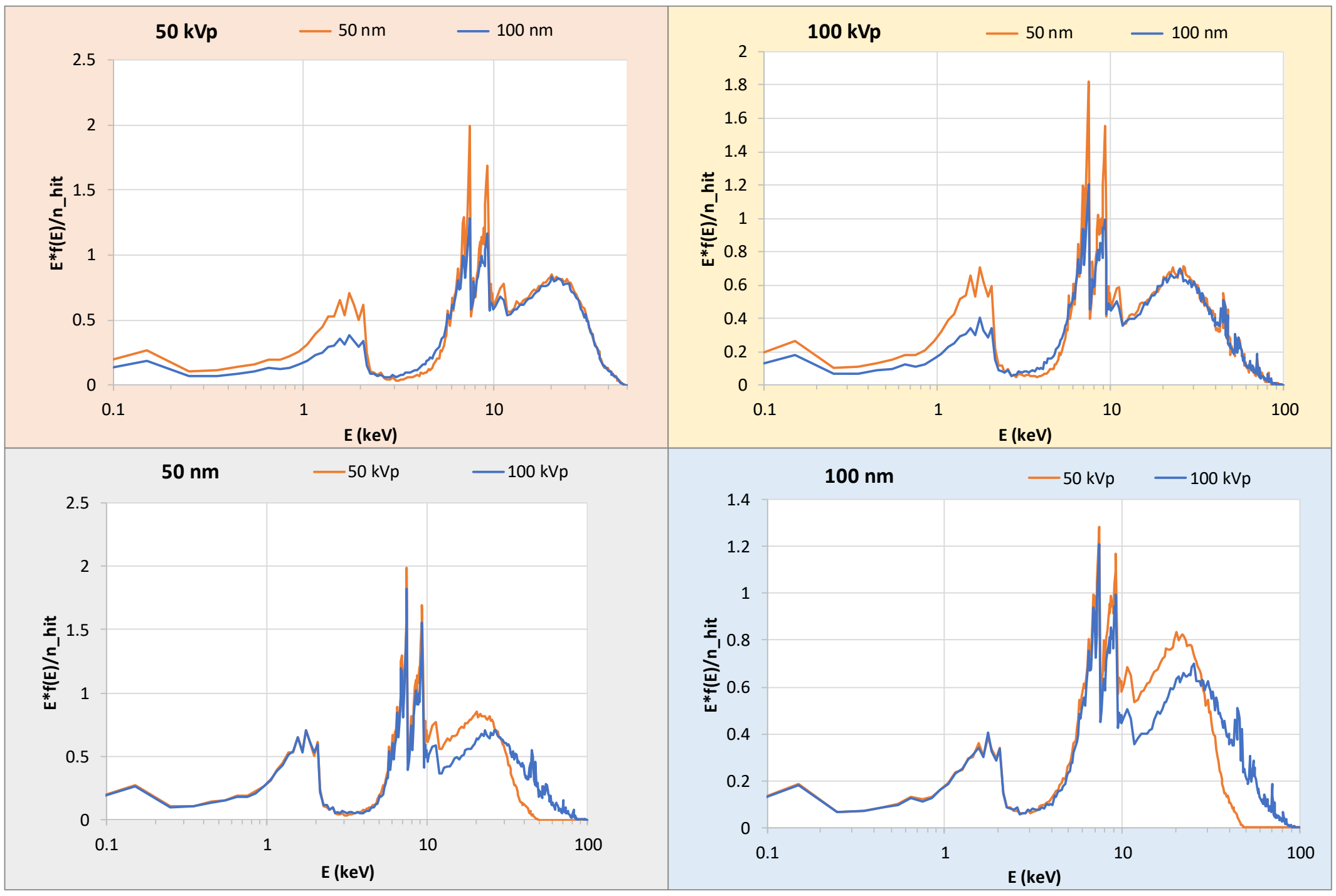

Figure 18: Electron energy flux per photon interaction in a GNP derived from the originally reported data of participant G4/DNA#1.

The correctly normalized original results from this participant are shown in Figure 18 and show a similar picture as was seen in Figure 5, i.e. agreement for energies above 10 keV for the same radiation quality and different GNP size and below 10 keV for the same GNP size and the two photon spectra.

The same processing error also occurred for the data of participants G4/DNA#3 and MCNP6 where normalization to the energy bin width was omitted in all four cases. This observation was published as a corrigendum (Li *et al.*, 2020b) in the same journal to original publication (Li *et al.*, 2020a) .

## 7.3 Incorrectly implemented photon energy spectrum

### *7.3.1 Participant PENELOPE#1*

The initially reported data of participant PENELOPE#1 failed the consistency checks for energy deposition and emitted electron spectra, where the values obtained for energy deposition for different GNP size and photon spectrum combinations appeared consistent with each other and also generally with the reported electron spectra, except for the 50 nm GNP irradiated with 50 kVp photons.

Figure 19 reveals that this latter observation can be easily explained by assuming that the data reported for the 50 nm GNP irradiated by 50 kVp photons was in reality for a 100 nm GNP. The two spectra shown in the left top panel coincide in relative spectral shape and differ in absolute scale by a factor of about 2.5, which is the ratio of the photon interaction probabilities for the two GNP sizes.

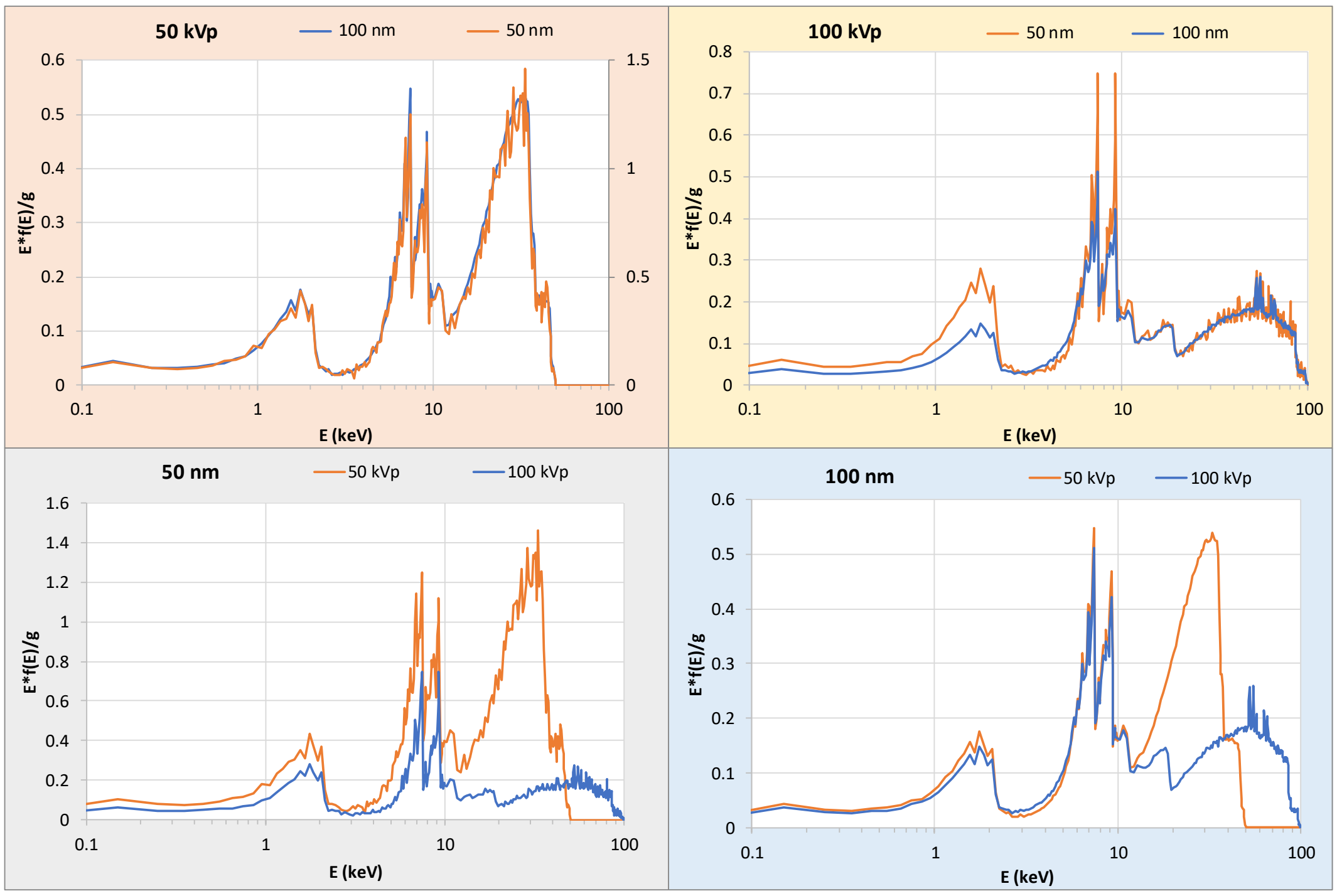

Figure 19: Electron flux spectra derived from the originally reported results of participant PENELOPE#1. In the left top panel, the right-hand side y-axis refers to the red curve.

Figure 20 shows the variation of the test quantity $C_w$ with sphere radius which shows very different functional shapes as compared to Figure 6. Figure 21 shows the radial dependence of excess energy deposited around a GNP experiencing a photon interaction, where the data are obviously not yet converging to zero. This indicates the presence of a larger fraction of long-range electrons, i.e., electrons of higher energy that is already evident from Figure 19.

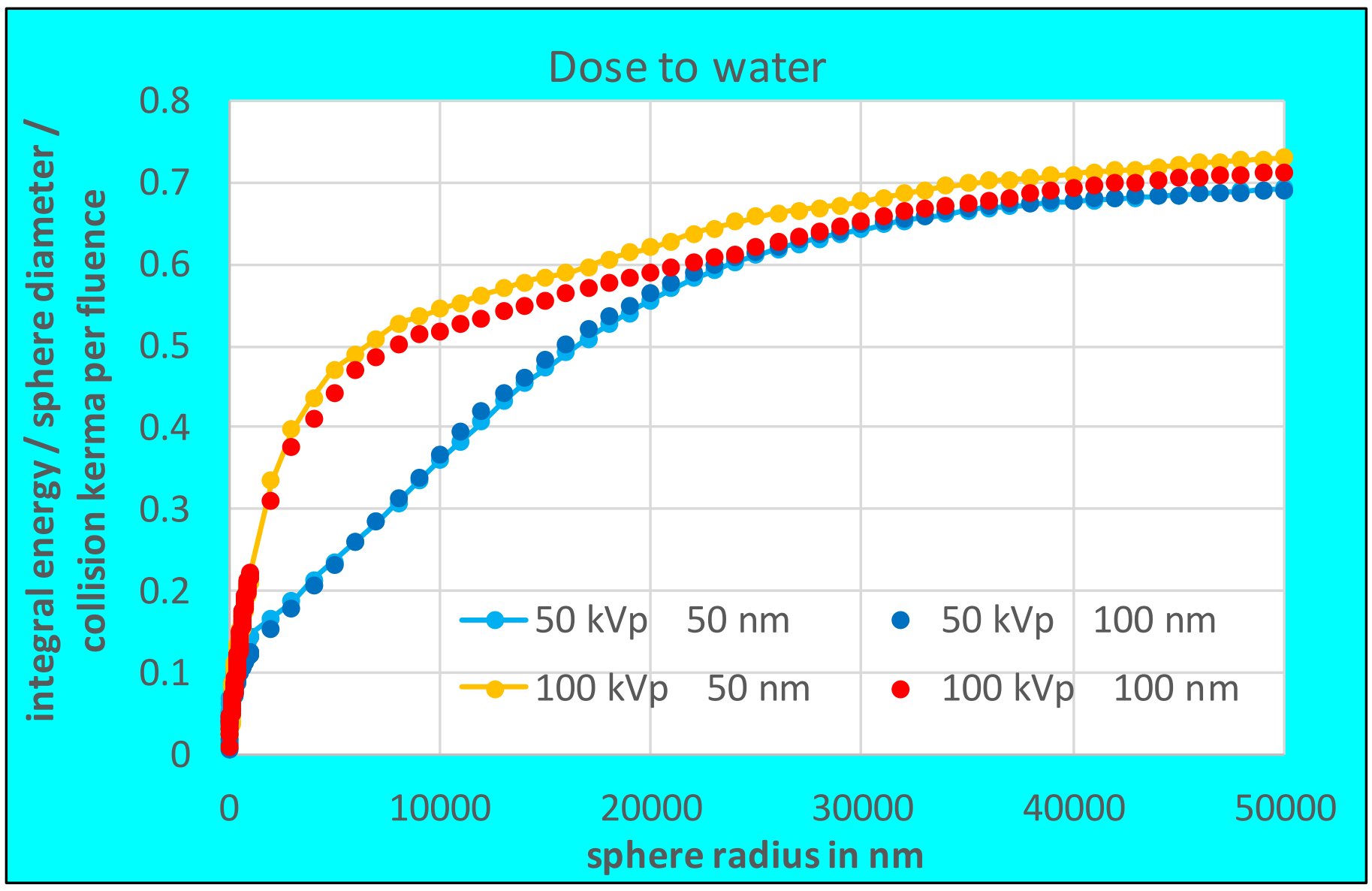

Figure 20: Ratio of the average dose within a water sphere (calculated from eq. (2)) and the collision kerma within the part of the sphere traversed by the primary photon beam (calculated from equation (3)) for the originally reported data of participant PENELOPE#1

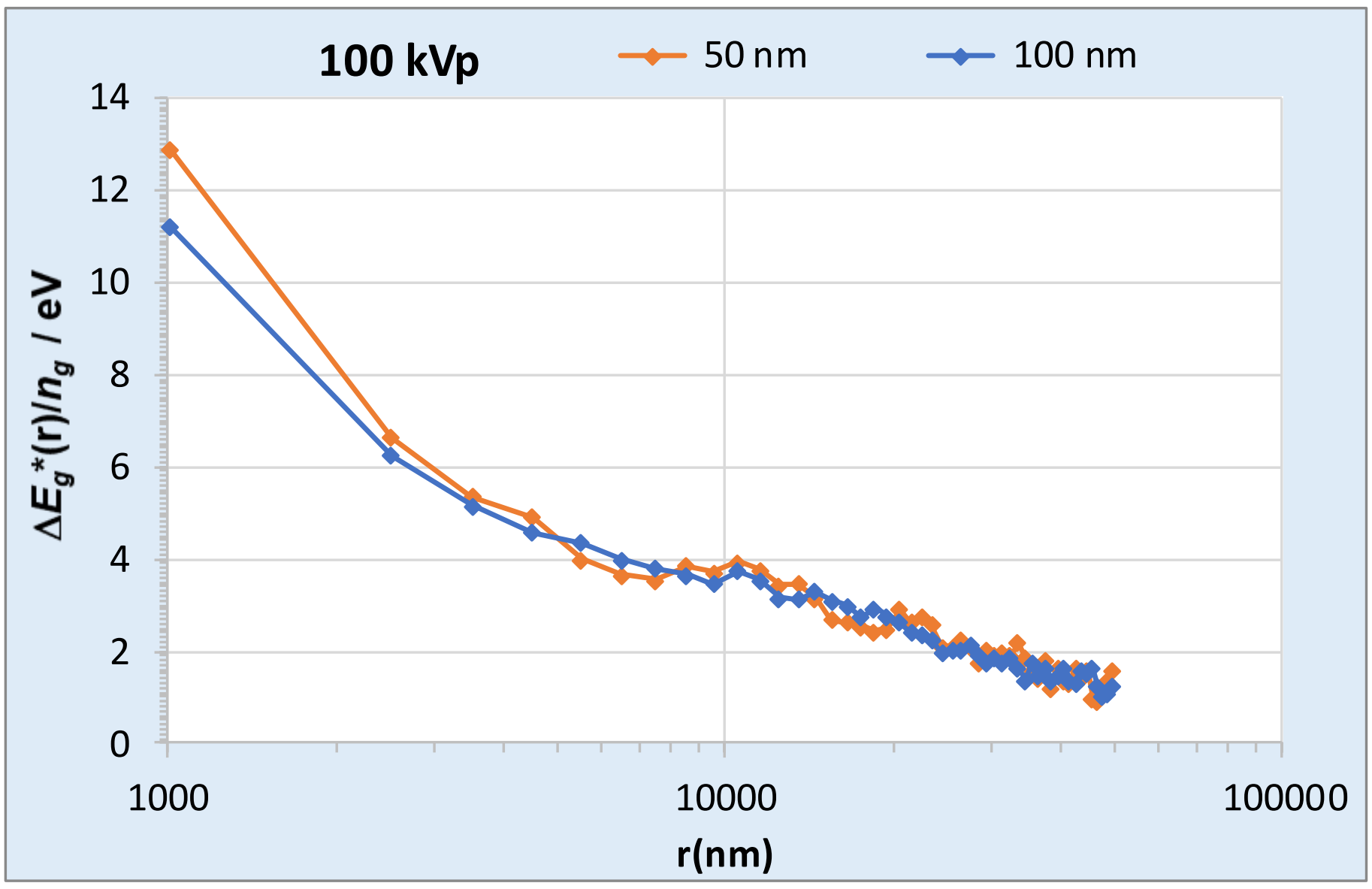

Figure 21: Excess imparted energy around a GNP where a photon interaction occurred for two GNP sizes and the 100 kVp photon spectrum. (Data from participant PENELOPE#1.)

This presumption is further illustrated in Figure 22 where the electron spectrum reported for the 100 nm GNP and 100 kVp photon spectrum is compared with energy fluence of the primary photon beam.

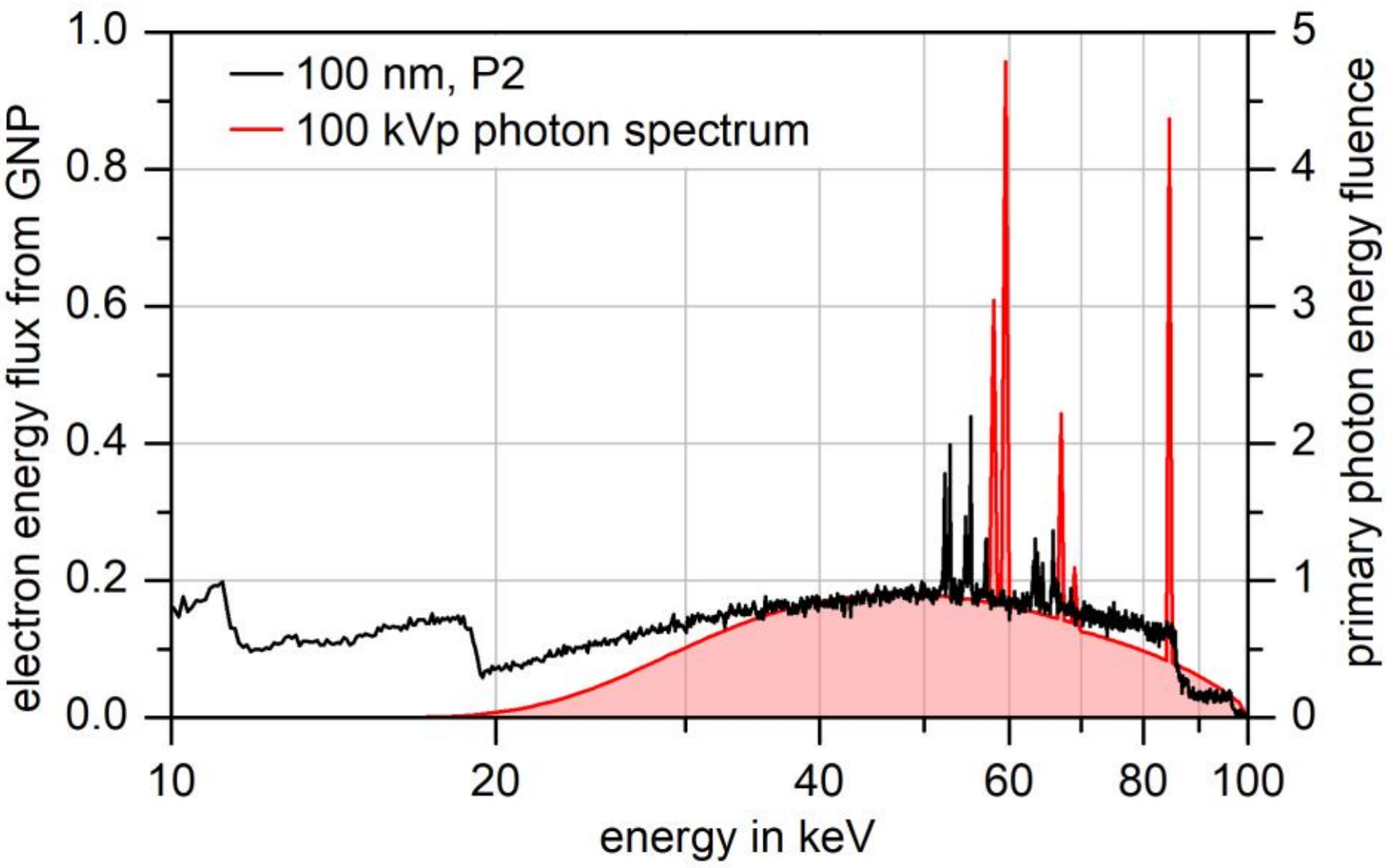

Figure 22: Comparison of the energy fluence of the primary 100 kVp photon spectrum and the energy spectrum of electrons leaving 100 nm GNP as reported by participant PENELOPE#1

The explanation for these observations was that in the input files for the simulations, the cumulative frequency rather than the frequency had been used for the photon energy spectrum, such that the reported results referred to a much harder X-ray spectrum. In consequence, participant PENELOPE#1 repeated all simulations (using the 2018 release of PENELOPE instead of the originally used 2011 release) and provided updated results. As the main programs provided with PENELOPE do not foresee the possibility of a circular source, the new simulations have been conducted with a square source of the same side length as the diameter of the circular source set in the exercise. This required a further correction (see next section).

### *7.3.2 Participant PENELOPE#2*

This participant originally only reported energy deposition information where the numbers given in Table 5 and Table 7 are similar to the values found for participant PENELOPE#1. The radial dependence of the reported dose enhancement was also quite similar. During the feedback loop it turned out that this participant had copied the simulation input files from participant PENELOPE#1 and used them for running simulations with a different release of PENELOPE). Therefore, the results are similarly compromised as the original results of PENELOPE#1.

Participant PENELOPE#2 did not provide new simulation results, so that with respect to the final outcome of the exercise, the results of participant PENELOPE#2 are considered as withdrawn.

## 7.4 Incorrectly implemented irradiation geometry

### *7.4.1 Participant PENELOPE#1*

As mentioned above, owing to limitations of the main programs provided with the PENELOPE distributions, the participant used a source geometry that was a square with the same side length as the circular source diameter defined in the exercise. Therefore, the revised simulation results hat to be multiplied with a fluence correction factor according to eq. (12) of $F_{\Phi} = 4/\pi$. In addition, corrected DER values were also determined according to eq. (13).

After the correction, the updated results passed the consistency checks in all cases except for the parameter $C_w$ and the 50 kVp spectrum (Table 13). In fact, all $C_w$ values are by about 3% lower than the results obtained for the data of participants using GEANT4 or derived codes. This may reflect difference in cross sections and the general uncertainty of interaction cross sections (Andreo *et al.*, 2012) .

Table 13: Results of the consistency checks for the revised data of participant PENELOPE#1.

| | 50 kVp photon spectrum | | 100 kVp photon spectrum | |
|---|---|---|---|---|
| | 50 nm GNP | 100 nm GNP | 50 nm GNP | 100 nm GNP |
| $C_w$ | 0.98 | 0.99 | 1.02 | 1.02 |
| $C_g$ | 0.86 | 0.83 | 0.87 | 0.81 |
| $C_e$ | 0.85 | 0.81 | 0.88 | 0.84 |

### *7.4.2 Participant TOPAS*

As suggested by the values in Table 7 and Table 8, the results of this participant for energy deposition and electron spectrum appeared consistent with each other for each combination of GNP size and photon spectrum. However, the values for different GNP size and photon spectra are inconsistent as there should be higher energy loss within the larger GNP.

Figure 23 shows the originally reported electron energy spectra, where the top two panels show clear differences in the energy range above 10 keV that would not be expected. Closer analysis values in Table 7 and Table 8 suggested that participant used a source size for which the radius was 10 nm larger than the GNP radius (instead of the diameter).

This presumption about the simulation set-up was confirmed by the participant, and an ensuing fluence correction factors per eq. (12) of $F_{\Phi} = (35/30)^2$ for the 50 nm GNP and $F_{\Phi} = (60/55)^2$ for the 100 nm GNP were applied to the energy deposition data and the electron spectra. The corrected electron spectra are shown in Figure 24 and show consistency between the different combinations of GNP size and photon spectrum.

In addition, corrected values for energy imparted and DER were also determined according to Eqs. (13) and(12)(15).

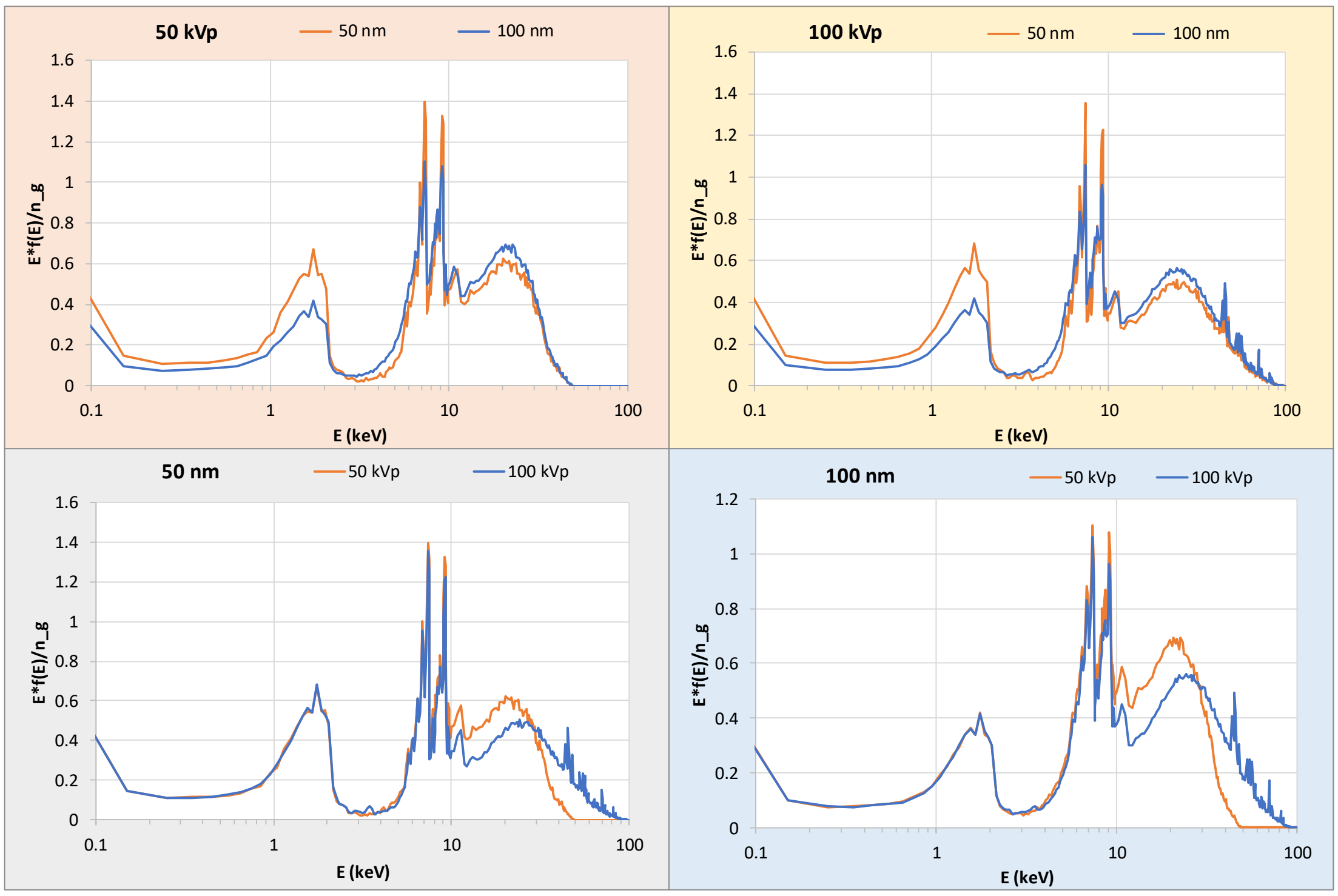

Figure 23: Comparison for the original data from participant TOPAS of the estimated energy flux of electrons emitted from a GNP that experienced a photon interaction for the two considered GNP sizes and the two photon energy spectra.

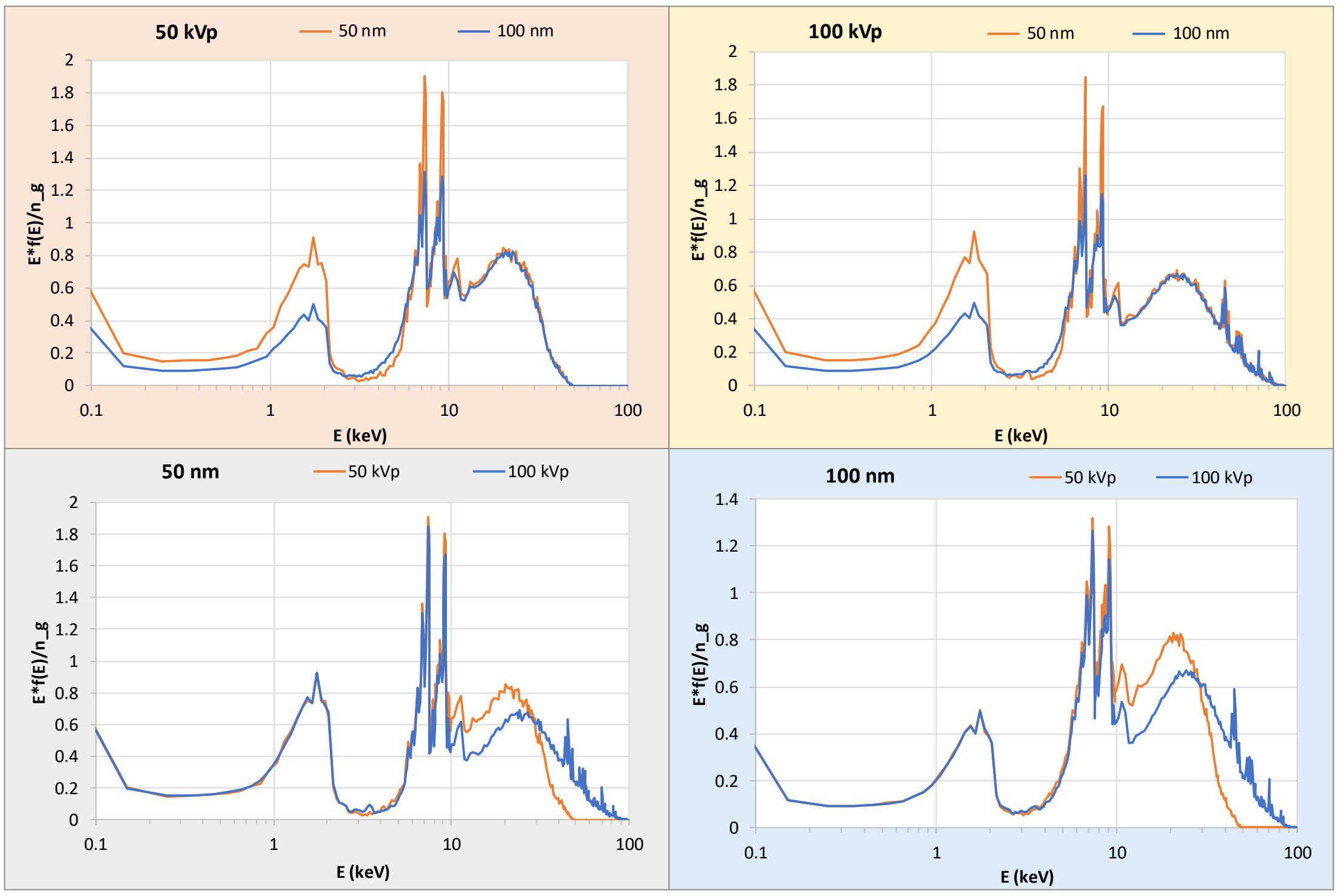

Figure 24: Comparison for the fluence-corrected data from participant TOPAS of the estimated energy flux of electrons emitted from a GNP that experienced a photon interaction for the two considered GNP sizes and the two photon energy spectra.

## 7.5 Incorrect emission angle range for scoring of emitted electrons

For participant MCNP6 all electron spectra failed the consistency checks, even after the normalization with the energy bin width of 50 eV had been done. Further interaction with the participant revealed that the electron spectra had only been recorded for close to normal emission from the surface. The participant performed additional simulations to estimate an overall scaling factor between the scored polar angle range and the full hemisphere that amounted to 11 (Li *et al.*, 2020b). The revised electron spectra (after bin-size normalization and this correction) passed the consistency tests.

The other remarkable observation is the low value for the correlation coefficient in Table 6, which is related to the fact that for this participant the reported data are generally exhibiting worse statistics than most of the other results.

## 7.6 Unresolved issues

### *7.6.1 Participant G4/DNA#2*

This participant only reported electron energy spectra for the 50 kVp photon spectrum. Figure 25 shows that both spectra fail the consistency checks, even after taking into account that the spectra were reported for equal binning of the logarithm of the electron energy. It can be seen from Figure 25 that the simulations are compromised by two problems: The spectrum for the 50 nm GNP is lacking signal in the energy range of the M-shell Auger lines between 1 keV and 3 keV. Both spectra shown in Figure 25 furthermore seem to have wrong energy positions of the L-shell Auger lines. The origin of these discrepancies was not fully explained by the participant (Li *et al.*, 2020b) .

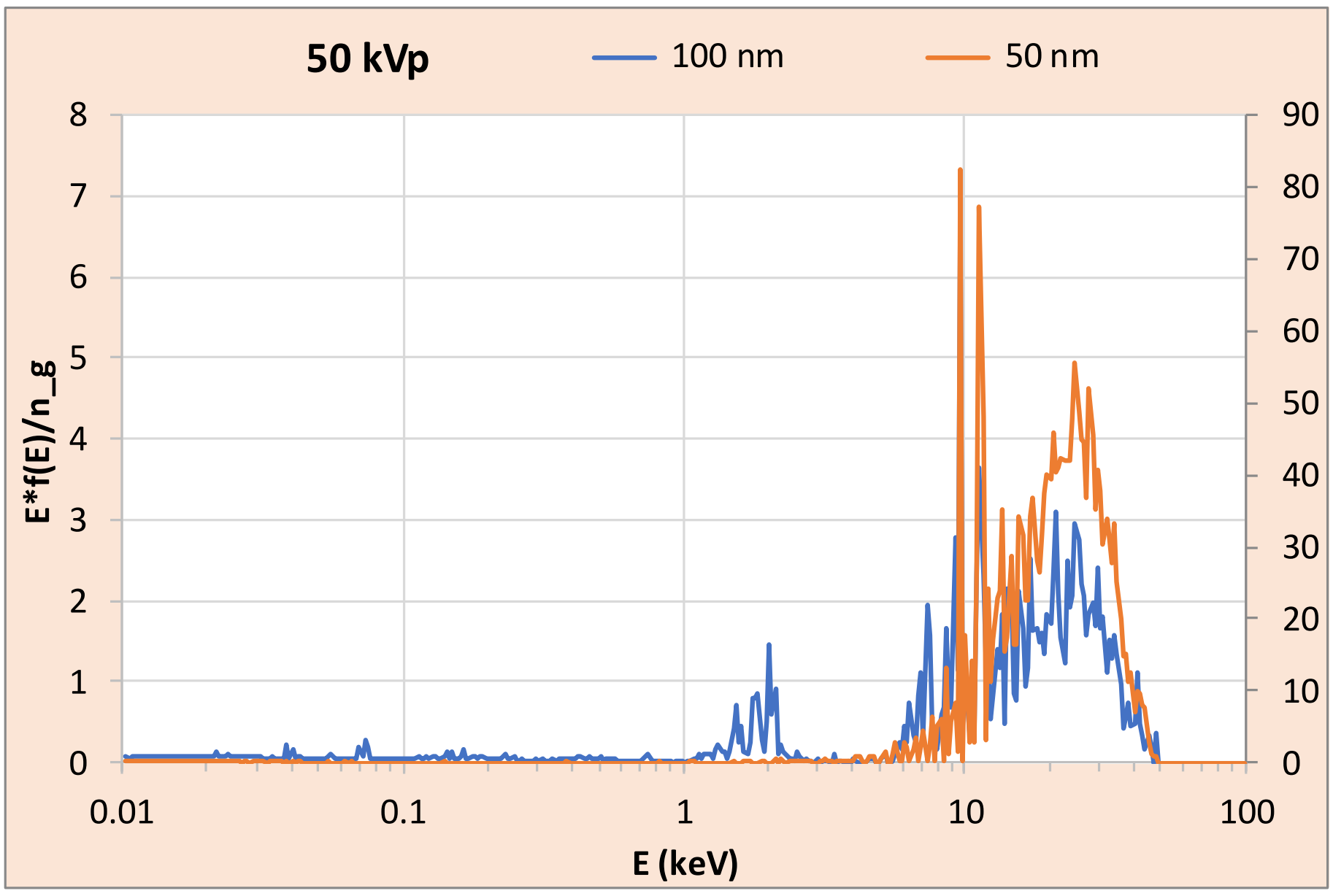

Figure 25: Electron energy flux per photon interaction in a GNP derived from the originally reported data of participant G4/DNA#2.

## *7.6.2 Participant G4/DNA#3*

For this participant, the reported results did not pass the consistency checks for energy conservation (see Table 7 and Table 8). From the ratio $C_e$ a lack of normalization to the energy bin width was suggested as the origin which was confirmed as being an error that occurred in processing the data for publication (Li *et al.*, 2020b) . Correcting this lacking normalization did not remove the failure of the electron spectra to pass the consistency tests. Furthermore, as is illustrated in Figure 26, there is also an issue regarding the internal consistency of the complete set of data from this participant.

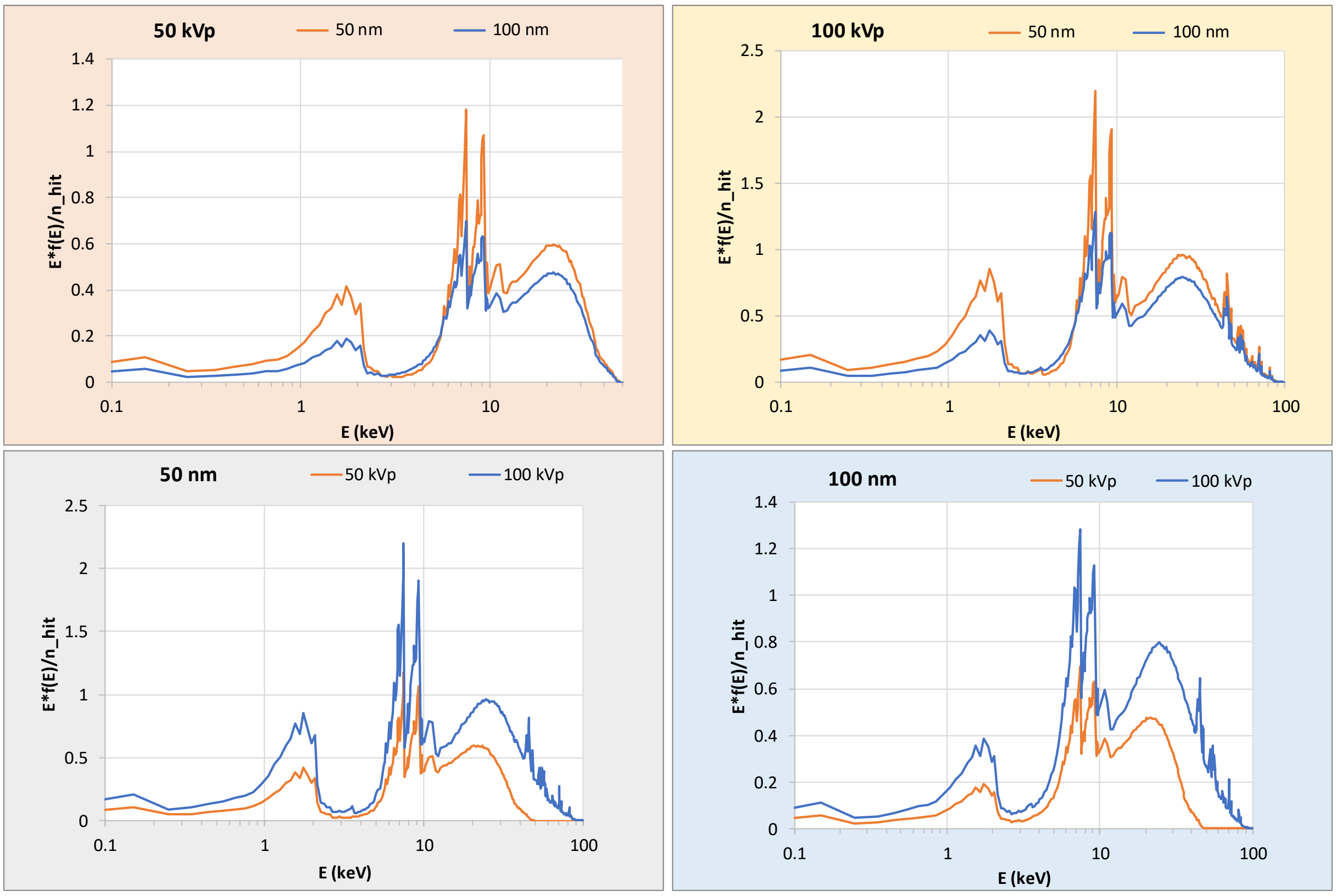

Figure 26: Electron spectra reported by participant G4/DNA#3 after normalization to the energy bin size. The data appear internally inconsistent as the differences for same photon spectrum in the energy range above 10 keV (cf. Figure 5) and for the same GNP size in the energy range below 10 keV (cf. Figure 4) are too pronounced.

Heuristic analysis showed that the discrepancies seen in Figure 26 as well as the values seen in Table 7 could be explained, if the simulation geometry was such that the beam cross section was identical to the cross section of the GNP rather than having a 10 nm larger diameter. Attempting a correction for such a scenario did also bring the test quantity $C_e$ to a value that would pass the consistency test for the 100 kVp spectrum while for the 50 kVp spectrum there remains a discrepancy suggesting the data to be a factor of 2 too low (Table 14).

Table 14: Results of the consistency checks if the data of G4/DNA#3 are corrected under the hypothesis that the photon beam diameter was equal to the GNP diameter.

| | 50 kVp photon spectrum | | 100 kVp photon spectrum | |
|---|---|---|---|---|
| | 50 nm GNP | 100 nm GNP | 50 nm GNP | 100 nm GNP |
| $C_g$ | 0.88 | 0.84 | 0.88 | 0.85 |
| $C_e$ | 0.44 | 0.41 | 0.89 | 0.86 |

Feedback of the participant who checked the simulation set-up, however, did not confirm aforementioned hypothesis. In consequence, the data of this participant continue to fail the consistency tests.

### *7.6.3 Participant PARTRAC*

For this participant the reported DERs reported in Li at al. (2020a, 2020b) consistently converge to values of about 1.2 at very large distances from the GNP. The asymptotic value of ratio $C_w$ in Table 5 is about 0.85 for all cases, whereas in Table 7 inconsistently enhanced values are seen for the 50 nm case.

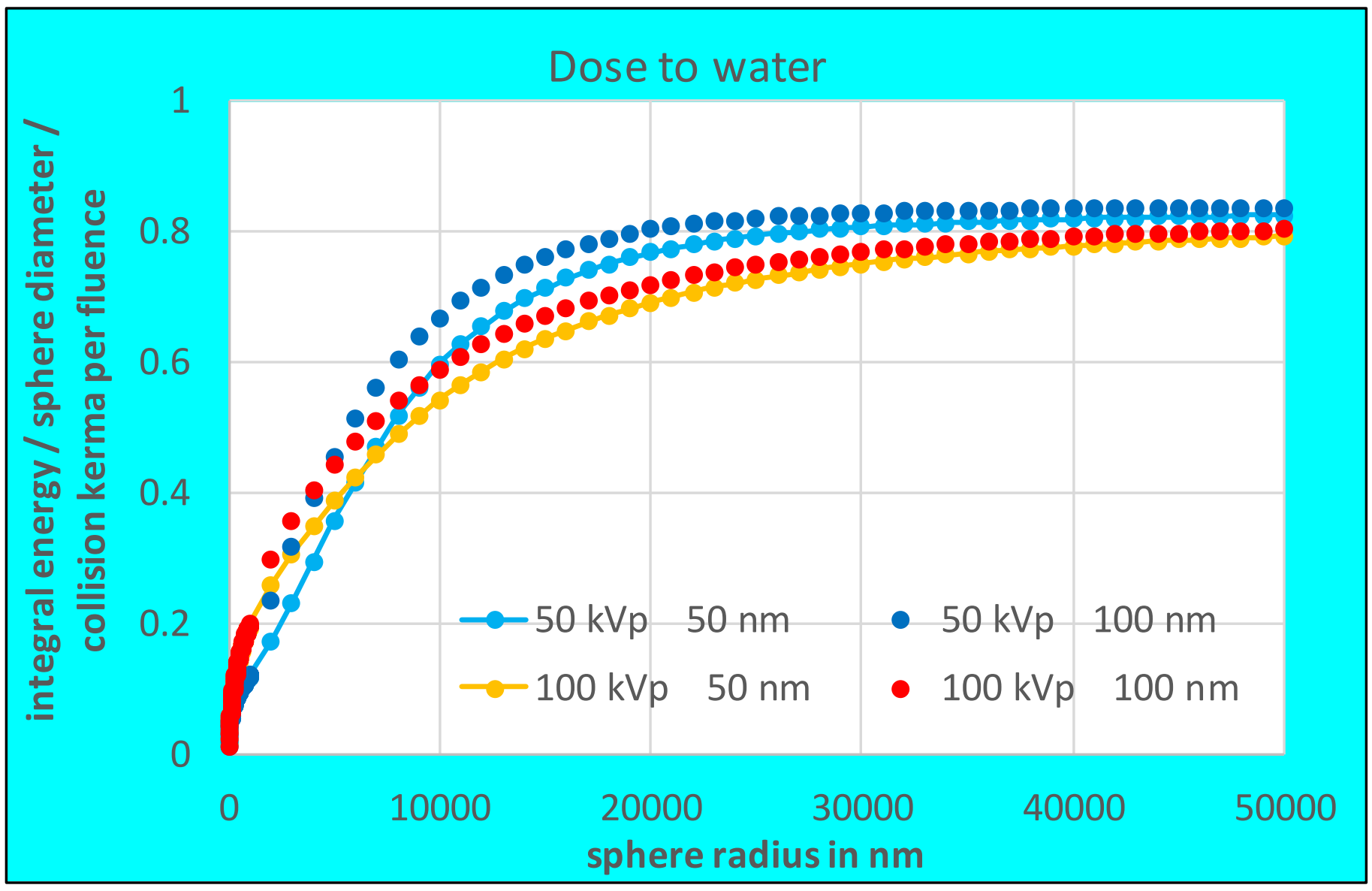

Figure 27: Radial dependence of the ratio $C_w$ derived from the data of participant PARTRAC.

The origin of these observed discrepancies could not be clarified, as they are not compatible with improper implemented geometry nor an overall normalization issue. Figure 27 shows that the ratio $C_w$ converges within the radial range considered in the scoring where the asymptotic value remains well below unity. It is remarkable, however, that the radial dependence varies between the two GNP sizes which is not to be expected.

In the course of the analysis, the hypothesis was tested that for some unknown reason, the simulation results for the GNP absent (and only these) are by a factor of 0.8 too low. If the consistency tests are performed after these simulation results have been multiplied by 1/0.8, the data of this participant would pass the consistency checks (see Table 15). In addition, also the deviation of the DER from unity would be removed.

Table 15: Outcome of the consistency checks for the data of participant PARTRAC, if a test-wise correction factor of 1.25 is applied to the simulation results without GNP.

| | 50 kVp photon spectrum | | 100 kVp photon spectrum | |
|---|---|---|---|---|
| | 50 nm GNP | 100 nm GNP | 50 nm GNP | 100 nm GNP |
| $C_w$ | 1.06 | 1.06 | 1.06 | 1.06 |
| $C_g$ | 0.87 | 0.71 | 0.88 | 0.77 |

These speculations were not confirmed by the participant and no alternative explanation was provided, so that the final data remain unchanged and continue to fail the consistency tests.

# 8. Supplementary Information

In this section, a description of the data files containing the averages and the estimated uncertainty bands derived from all data that finally passed the consistency checks (see Sections 4.2 and 4.3 for details). These data are meant to serve as reference solutions for simulations that newcomer to the field may want to attempt for training. It is understood that they apply only to the geometry, photon energy spectra, and tallies as described in the exercise definition (see Section 2).

The data files are contained in a zip-archive named EURADOS_Single_GNP-exercise.zip (doi: https://doi.org/XXXX-YYYY/ZZZZ).

## 8.1 Electron spectra

The electron spectra are provided in four Ascii data files named as follows:

- Electron_spectra_50kVp_50nm.csv
- Electron_spectra_50kVp_100nm.csv
- Electron_spectra_100kVp_50nm.csv
- Electron_spectra_100kVp_100nm.csv

which contain a short header (lines starting with a "#") and then four data columns separated by commas. The meaning of the data columns is as follows:

1. Lower boundary of the electron energy bin in eV
2. Center of the electron energy bin in eV
3. Mean frequency density of electrons in $eV^{-1}$ for a fluence of 1 photon per source area
4. Uncertainty of the frequency density in $eV^{-1}$ (estimated as two standard deviations)

## 8.2 Energy deposition in water shells

The energy deposition in water shells with and without GNP are provided as eight Ascii data files named as follows:

- Energy_deposition_with_GNP_50kVp_50nm.csv
- Energy_deposition_without_GNP_50kVp_50nm.csv
- Energy_deposition_with_GNP_50kVp_100nm.csv
- Energy_deposition_without_GNP_50kVp_100nm.csv
- Energy_deposition_with_GNP_100kVp_50nm.csv
- Energy_deposition_without_GNP_100kVp_50nm.csv
- Energy_deposition_with_GNP_100kVp_100nm.csv
- Energy_deposition_without_GNP_100kVp_100nm.csv

which contain a short header (lines starting with a "#") and then four data columns separated by commas. The meaning of the data columns is as follows:

1. Lower boundary of the bin for the distance from the GNP surface in nm
2. Lower boundary of the bin for the radial distance from the GNP center in nm
3. Energy deposited in the spherical shell in eV for a fluence of 1 photon per source area
4. Uncertainty of the energy in eV deposited in the spherical shell (two standard deviations)